\documentclass[aps,prb,twocolumn,floatfix,superscriptaddress,showpacs]{revtex4-1}
\usepackage{times}
\usepackage{epsfig}
\usepackage{amsmath}
\usepackage{amssymb}
\usepackage{graphicx}
\usepackage{dsfont}
\usepackage[colorlinks=true,citecolor=red,linkcolor=blue]{hyperref}
\newcommand{\tsh}{t_\mathrm{h}}
\newcommand{\qbc}{\mathrm{QBC}}
\newcommand{\bfk}{\mathbf{k}}
\newcommand{\bfa}{\mathbf{a}}
\newcommand{\bfb}{\mathbf{b}}
\newcommand{\bfA}{\mathbf{A}}

\newcommand{\bfka}{\mathbf{k}\cdot\mathbf{a}}
\begin{document}

\title{Quadratic band touching points and flat bands in two-dimensional topological Floquet systems}
\author{Liang Du}
\thanks{Liang Du and Xiaoting Zhou contributed equally to this work.}
\affiliation{Department of Physics, The University of Texas at Austin, Austin, TX 78712, USA}
\author{Xiaoting Zhou}
\thanks{Liang Du and Xiaoting Zhou contributed equally to this work.}
\affiliation{Department of Physics, The University of Texas at Austin, Austin, TX 78712, USA}
\author{Gregory A. Fiete}
\affiliation{Department of Physics, The University of Texas at Austin, Austin, TX 78712, USA}

\begin{abstract}
In this work we theoretically study, using Floquet-Bloch theory, the influence of circularly and linearly polarized light on two-dimensional band structures with Dirac and quadratic band touching points, and flat bands, taking the nearest neighbor hopping model on the kagome lattice as an example.  We find circularly polarized light can invert the ordering of this three band model, while leaving the flat-band dispersionless.   We find a small gap is also opened at the quadratic band touching point by 2-photon and higher order processes.  By contrast, linearly polarized light splits the quadratic band touching point (into two Dirac points) by an amount that depends only on the amplitude and polarization direction of the light, independent of the frequency, and generally renders dispersion to the flat band. The splitting is perpendicular to the direction of the polarization of the light.  We derive an effective low-energy theory that captures these key results.  Finally, we compute the frequency dependence of the optical conductivity for this 3-band model and analyze the various interband contributions of the Floquet modes.  Our results suggest strategies for optically controlling band structure and interaction strength in real systems.
\end{abstract}
\date{\today}
\maketitle

\section{INTRODUCTION}
\label{sec:intro}

The past decade has seen dramatic advances in understanding the topological properties of the band structure of quantum many-particle systems. \cite{Moore:nat10,Hasan:rmp10,Qi:rmp11,Ando:jpsj13}  When inter-particle interactions are brought into the picture, the phenomenology is even richer. \cite{Maciejko:np15,Stern:arcmp15,witczak-krempa2014,mesaros2013,Chen:prb13}  Certain features of two-dimensional band structures are known to have particular stability conditions with respect to inter-particle interactions.\cite{Sun:prl09}  For example, Dirac points are perturbatively stable to interactions, requiring a critical interaction strength to open a gap,\cite{Meng:nat10,Hohenadler:prl11,Yu:prl11} which is important for the low-energy properties of single-layer graphene.\cite{Neto:rmp09}   By contrast, quadratic band touching points in two dimensions are known to be perturbatively unstable to interactions,\cite{Sun:prl09} and numerous discussions in the literature of ``interaction-driven" topological states have appeared (including in more general, higher dimensional, contexts).\cite{Wen:prb10,Weeks:prb10a,Zhang:prb09,Raghu:prl08,Ruegg:prl12,Yang:prb11a,Ruegg11_2,Wang:epl12,Yoshida:prb13,Go:prl12,Maciejko:prl14,Pesin:np10,Kargarian:prb10,Shitade:prl09,Dzero:prl10,Budich:prb12,Budich:prb13}  In addition, flat bands in any spatial dimension correspond to localized states which generally tend to have enhanced interaction effects because the energy scale of interactions, no matter how small, will always dominate the kinetic energy scale in a flat band.\cite{Wang:prl11}  One particularly active research topic in this area has been the study of fermionic lattice models where the lowest band is nearly flat, has a finite Chern number, and is separated by a gap large compared to the band width of the lowest band.\cite{Sheng11,Sun:prl11,Neupert:prl11,Tang:prl11}  If this band is partially filled, (e.g., 1/3 filled), fractional quantum Hall states may arise in the lattice model.\cite{Wu_CI:prb12,Liu:prb13,Parameswaran2013816,Qi11,Neupert:prb11,Regnault:prx11,Kourtis:prl14}  

Another direction the study of topological phases has taken in recent years is the non-equilibrium generation of interesting band structures under the influence of a periodic drive.\cite{Lindner:np11}  At the non-interacting level, dramatic changes in the band structure can occur, including a change from a non-topological band structure to a topological one.\cite{Kitagawa:prb10,Rudner:prx13,Katan:prl13,Lindner:prb13,Dora:prl12,Inoue:prl10,Cayssol:pss13,Kitagawa:prb11,Iadecola:prl13,Ezawa:prl13,Kemper:prb13,Rechtsman:nat13}  Two commonly discussed physical scenarios for periodically driven systems include periodic changes in the laser fields that establish the optical lattice potential for cold atom systems,\cite{Jotzu:nat14,Bilitewski:pra15} and solid state systems that are driven by a monochromatic laser field.\cite{Fregoso:prb13,sentef2015theory,Wang:sci13,Mahmood:np16,PhysRevB.91.241404,PhysRevA.92.023624, PhysRevA.91.043625, PhysRevB.89.121401}  When inter-particle interactions are included in such Floquet-Bloch systems, energy is typically absorbed from the periodic drive.\cite{Kim:pre14}  If the many-particle system is closed, it will usually end up at infinite temperature, unless the system is sufficiently disordered for many-body localization\cite{Ponte:prl15,Lazarides:prl15,Genske:pra15} or some other non-generic state to occur.  If the system is coupled to a bath, it is possible for a ``balance" to be established where the average energy (over a drive period) absorbed by the system from the drive can be released to the bath and a non-equilibrium steady state established.\cite{Hossein:prb14,Hossein_2:prb16,Iadecola:prb15,Iadecola_2:prb15,Seetharam:prx15,Shirai:pre15}  Even when such a non-equilibrium steady state is established, the distribution of particles in their orbitals is generally non-thermal, and non-generic, i.e., depends on the properties of the system in the absence of the drive, the drive itself, and the details of the bath.\cite{Hossein:prb14,Hossein_2:prb16,Iadecola:prb15,Iadecola_2:prb15,Seetharam:prx15,Shirai:pre15}  In real systems, such experimental conditions are ideal for potentially realizing novel states of matter that might be absent from the equilibrium phase diagram of the system.\cite{Bukov:prl15}

In this work, we focus on the features of the underlying band structure.  To date, much of the theoretical work on Floquet-Bloch systems has studied two-band models, because they are the simplest case for which a trivial band structure can be ``reorganized" into a topological band structure.\cite{PhysRevB.91.241404,PhysRevA.92.023624, PhysRevA.91.043625,Lindner:np11,Hossein:prb14,Aditiprb91-2015,Aditiprb92-2015,Hossein:prb16,Hossein_2:prb16}  In fact, the physics of ``band inversions" much discussed in the literature of time-reversal invariant topological insulators\cite{Hasan:rmp10,Qi:rmp11} has strong analogs in Floquet-Bloch systems.  Though, the bulk-boundary correspondence is much more complicated in the Floquet-Bloch systems than in the equilibrium systems,\cite{Rudner:prx13} and transport measurements on two-dimensional Floquet-Bloch systems with topological band structures are also much less straight-forward to interpret.\cite{Torres:prl14,Kundu:prl14,Farrell:prb16}   

On a technical level, the Floquet-Bloch theory is often studied as an expansion about the high-frequency limit.\cite{D?Alessio201319}  It is therefore not clear {\it a priori} that it is sensible for a multi-band system to be reduced to an effective, low-energy two-band model about the Fermi energy.  Possible resonances between these low-energy bands, and higher energy bands in a real system might have a significant impact on the Floquet band structures.  As a step towards understanding this physics, we study a two-dimensional three-band fermion model that also includes a flat band, a quadratic band touching point, and a Dirac point (all in the absence of the drive) to investigate how these features are impacted by the drive from normally incident laser fields of different intensities, frequencies, and polarizations.  We find a rich phenomenology that in some cases mimics the changes that may be induced by interactions in equilibrium band structures,\cite{Sun:prl09} but also has many unique features of its own.  This opens the possibility that one may be able to optically engineer topological states, including those that might have non-trivial inter-particle interactions at their core.\cite{Grushin:prl14} 

Our paper is organized as follows.  In Sec. \ref{sec:Ham}, we describe the three-band kagome lattice Hamiltonian we study, and in Secs.\ref{sec:drive},\ref{sec:Floquet} we discuss the influence of a monochromatic laser field of different polarizations, intensities, and frequencies on the Hamiltonian.  In Sec.\ref{sec:low_energy} we obtain an effective low-energy theory about the quadratic and Dirac band touching points that describes the opening of gaps in the presence of a laser field in those systems.  In Sec.\ref{sec:sigma}, we compute the finite-frequency optical conductivity of the model for different laser parameters, and in Sec.\ref{sec:conclusions} we summarize the main conclusions of this work.

\section{Model Hamiltonian and Band Structure}
\label{sec:Ham}

\begin{figure}[h]
\includegraphics[width=1.0\linewidth]{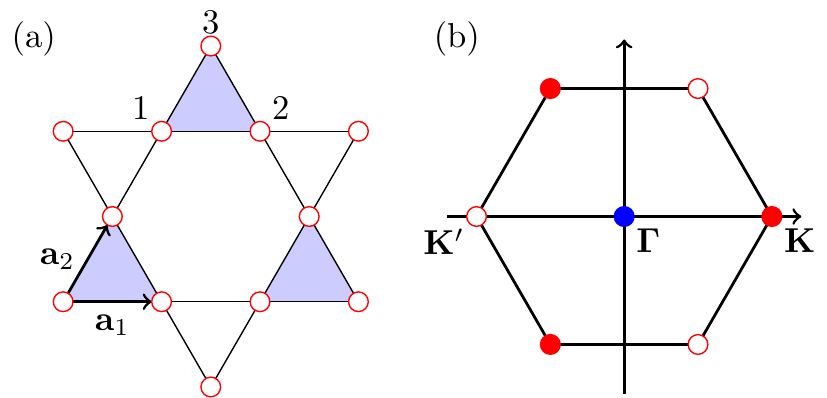}  %
\caption{(Color online) (a) Kagome lattice with nearest neighbor vectors ${\bfa_1,\bfa_2}$ labelled, and the three sites in the unit cell labelled.  (b) First Brillouin zone of the underlying triangular Bravais lattice.  In the nearest-neighbor hopping model, Eq.\eqref{eq:htbr}, Dirac points occur at the ${\bf K}$ and ${\bf K'}$ points, and a quadratic band touching point at the ${\bf \Gamma}$ point.  All red (gray) circles are equivalent to ${\bf K}$ and all open circles are equivalent to  ${\bf K'}$.
}
\label{fig:kagome_lattice}
\end{figure}
The three-band model we study is based on the nearest neighbor hopping model, Eq.\eqref{eq:htbr}, defined on the kagome lattice, a two-dimensional corner sharing network of triangles shown in Fig.\ref{fig:kagome_lattice}(a). The high symmetry points in the first Brillouin are indicated in  Fig.\ref{fig:kagome_lattice}(b), with Dirac points at the ${\bf K}$ and ${\bf K'}$ points, and a quadratic band touching point at the ${\bf \Gamma}$ point.

We study the tight-binding Hamiltonian on the kagome lattice,
\begin{equation}\label{eq:htbr}
     H = -t_{\text h} \sum_{\langle ij \rangle, \sigma} c_{i\sigma}^{\dagger} c_{j\sigma},
\end{equation}
where $t_{\text h}$ is the isotropic hopping integral between nearest neighbours, 
$c_{i\sigma}^\dagger$ ($c_{j\sigma}$) creates (annihilate) an electron with spin $\sigma$ 
on site $i$ ($j$) of the kagome lattice, and $\langle ij \rangle$ denotes nearest neighbors.  A related model has also been studied in the context of topological insulators\cite{Guo:prb09,Wen:prb10} and Chern insulators.\cite{Tang:prl11}

Fourier transforming to momentum space, the Hamiltonian becomes,
$H = \sum_{{\bf k}, \sigma} \psi_{{\bf k}\sigma}^\dagger {H}_{\bf k} \psi_{{\bf k}\sigma}$ with 
$\psi_{{\bf k}\sigma}=(a_{{\bf k}\sigma}, b_{{\bf k}\sigma}, c_{{\bf k}\sigma})^T$,
%>>> time independent tight-binding Hamiltonian in momentum space
\begin{equation}\label{eq:htbk}
    {H}_{\mathbf{k}} = -2 t_{\text h}
    \begin{pmatrix}
        0 & \cos(\mathbf{k} \cdot {\mathbf a}_{1}) & \cos(\mathbf{k} \cdot \mathbf{a}_{2})\\
        \cos(\mathbf{k} \cdot \mathbf{a}_{1}) & 0 & \cos(\mathbf{k} \cdot \mathbf{a}_{3})\\
        \cos(\mathbf{k} \cdot \mathbf{a}_{2}) & \cos(\mathbf{k} \cdot \mathbf{a}_{3}) & 0
    \end{pmatrix},
\end{equation}
%>>> symbols used in the equation
where $a_{\mathbf{k}\sigma}$, $b_{\mathbf{k}\sigma}$ and 
$c_{\mathbf{k}\sigma}$ defines annihilate operators on the three basis sites in the triangular unit cell shown in Fig.\ref{fig:kagome_lattice}(a). 
By setting the distance between nearest neighbors to be $1$, 
the nearest neighbor vectors are $\mathbf{a}_{1}=(1,0)$, $\mathbf{a}_{2}=(1/2,\sqrt{3}/2)$, and
$\mathbf{a}_{3}=\mathbf{a}_{2}-\mathbf{a}_{1} = (-1/2,\sqrt{3}/2)$. The translation verctors are $2 \mathbf{a}_1$ and $2 \mathbf{a}_2$.
The reciprocal-lattice primative vectors can be chosen as $\bfb_1 = \pi(1, -1/\sqrt{3})$, $\bfb_2 = \pi(0, 2/\sqrt{3})$.

%Note here the Hamiltonian above is not invariant under translation of integer combinations of $\bfb_1$ and $\bfb_2$. 
%Extra gauge transformations $b_{\bfk,\sigma} \rightarrow b_{\bfk\sigma}exp(-i\bfk\cdot\bfa_2)$, 
%$c_{\bfk\sigma} \rightarrow c_{\bfk\sigma}exp(-i\bfk\cdot\bfa_3)$ are needed to recover the translational symmetry in momentum space.

\begin{figure}[t]
\begin{center}
\includegraphics[width=4cm]{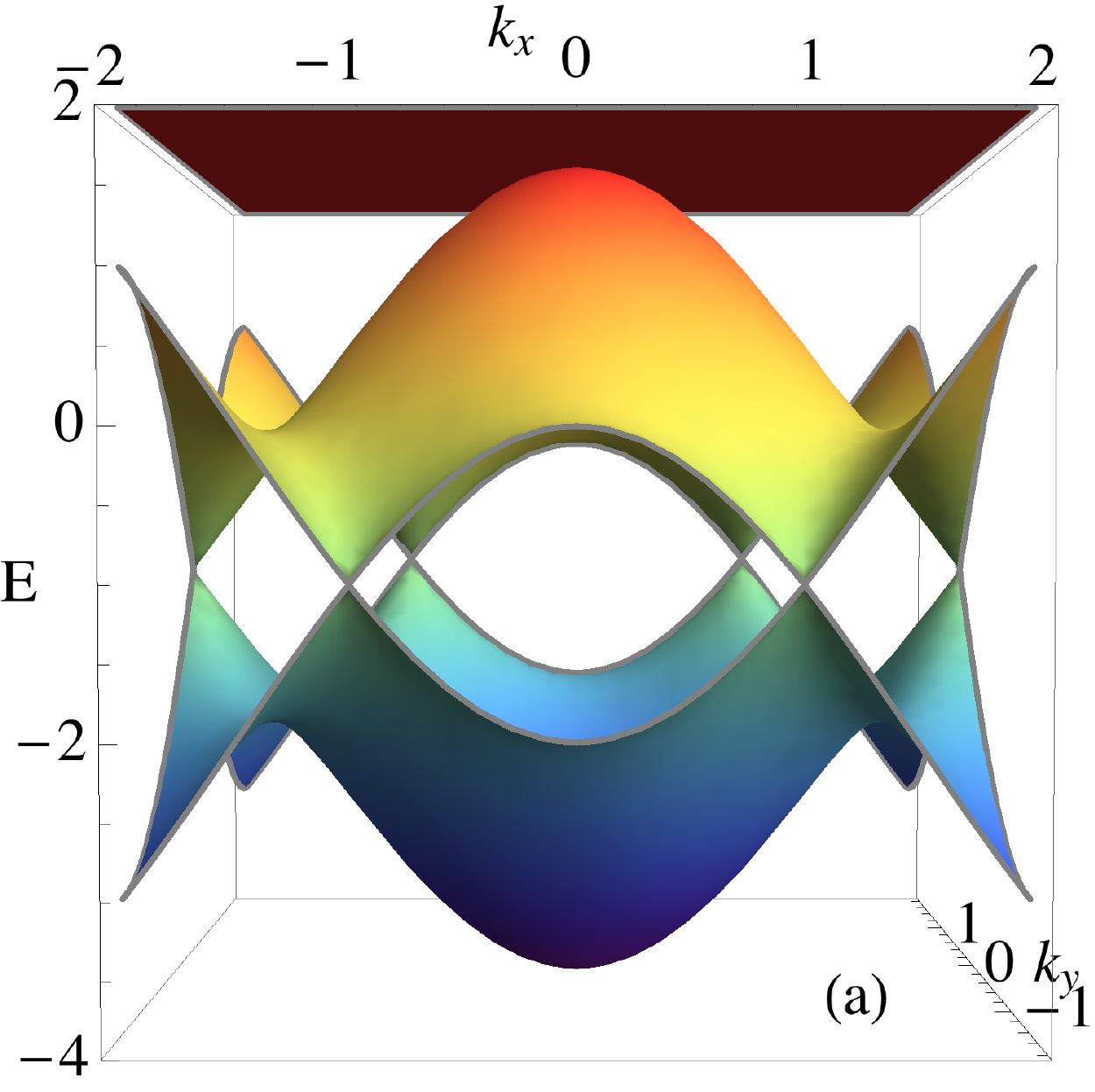}  %
\includegraphics[width=4cm]{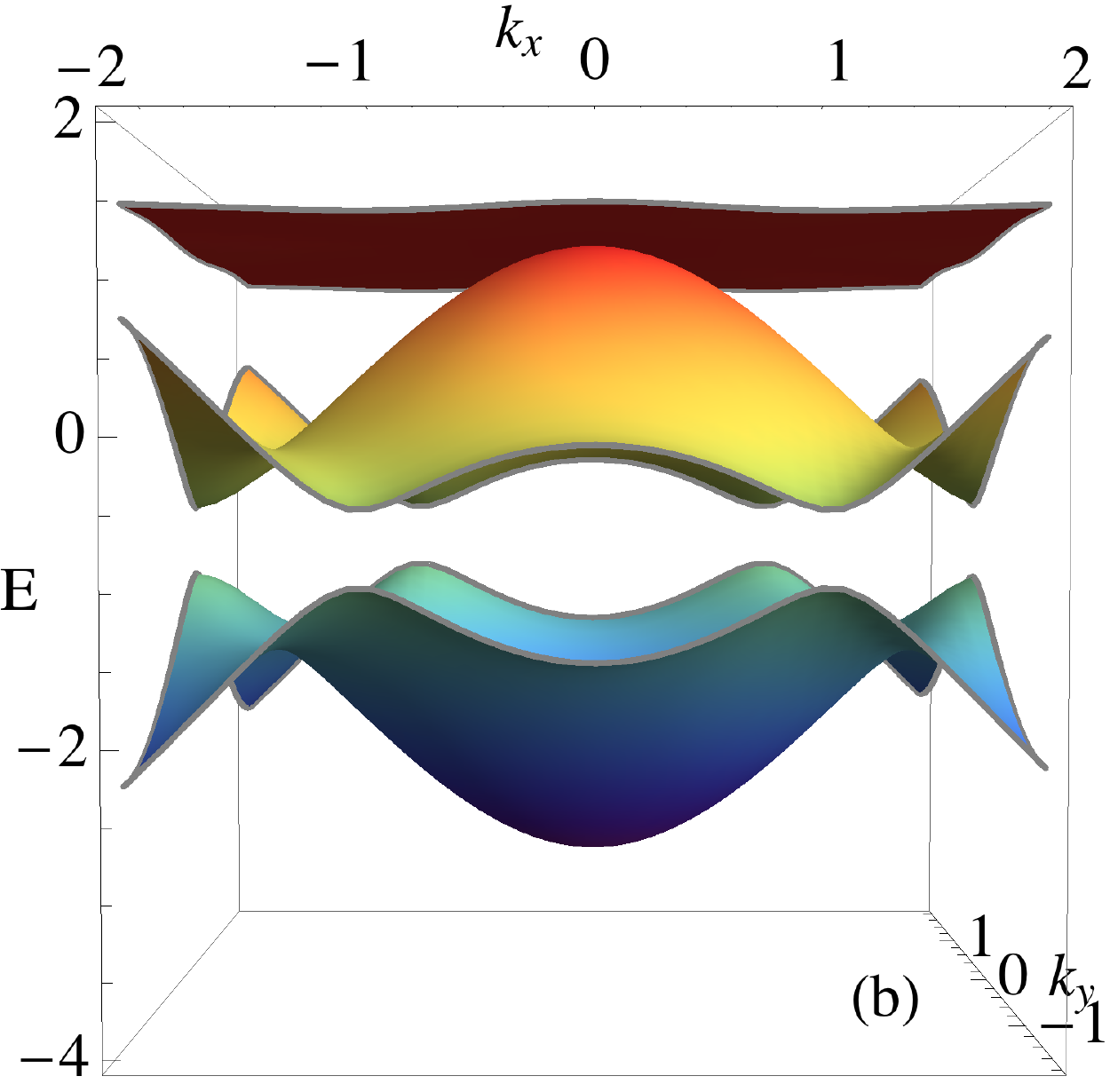}  %
\includegraphics[width=4cm]{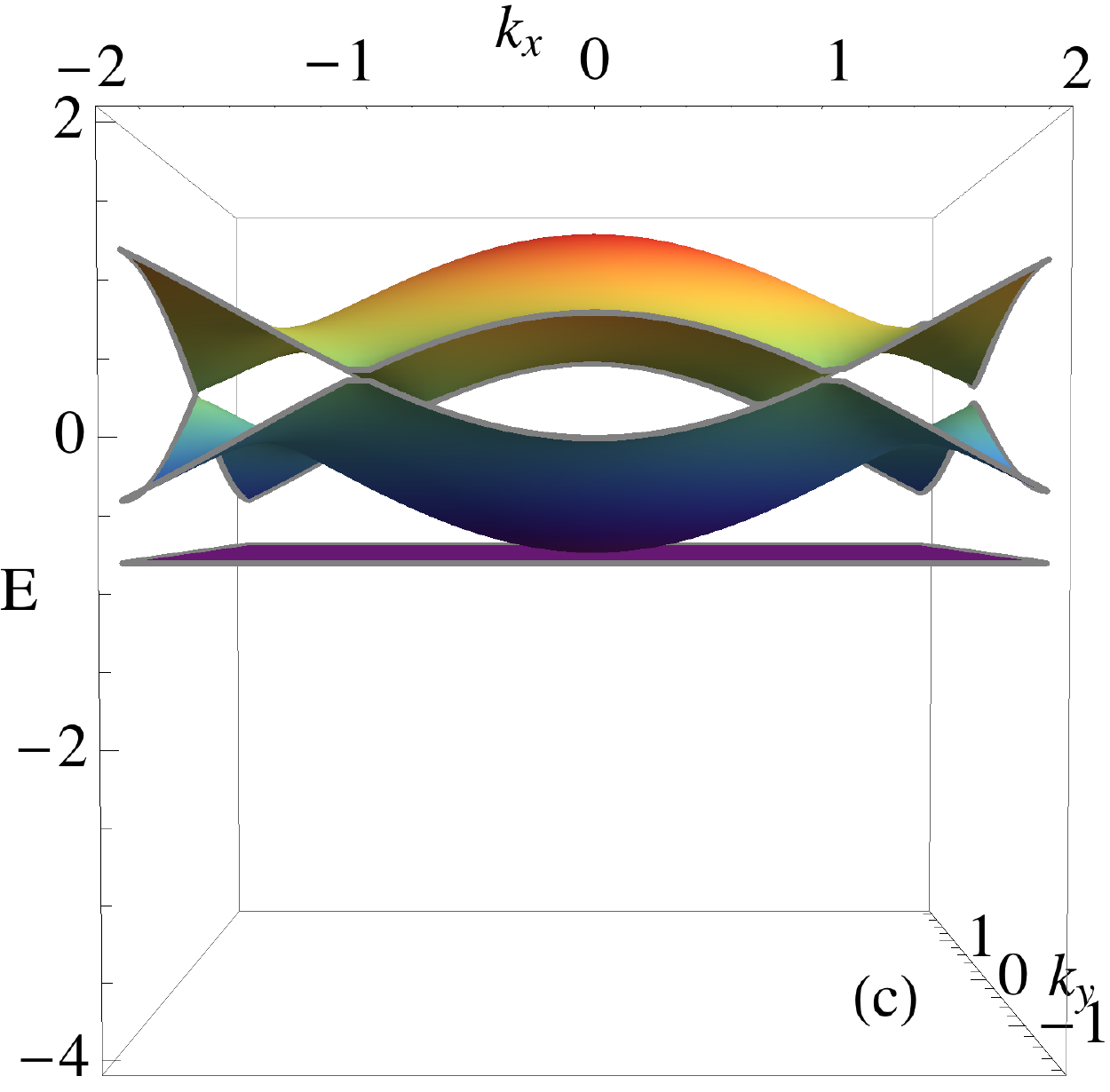}  %
\includegraphics[width=4cm]{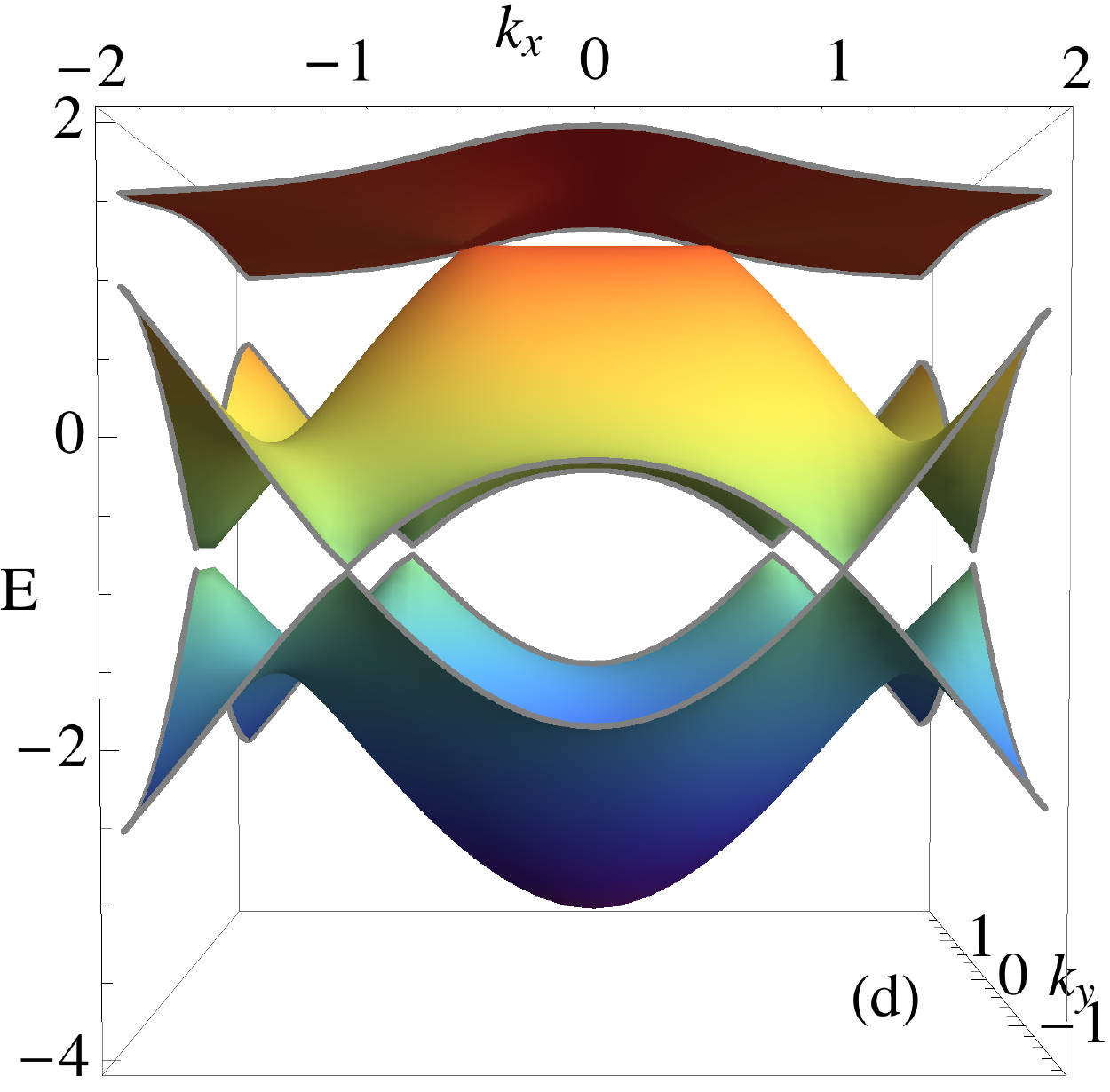}  %
\end{center}
\caption{(Color online) The Floquet-Bloch band structure of the kagome lattice embedded in a normally incident polarized light. 
The frequency of the pump light is $\Omega = 6\tsh$. (a) The band structure in equilibrium, without any incident light, given by Eq.\eqref{eq:htbr}; 
(b) The Floquet-Bloch band structure with circularly polarized light $A_0=1.0$; 
(c) The Floquet-Bloch band structure with circularly polarized light $A_0=3.8$, 
(d) The Floquet-Bloch band structure with linearly polarized light $A_0=1.0$. 
}
\label{fig:floquet-band-3d}
\end{figure}

\begin{figure}[t]
\begin{center}
\includegraphics[width=4cm]{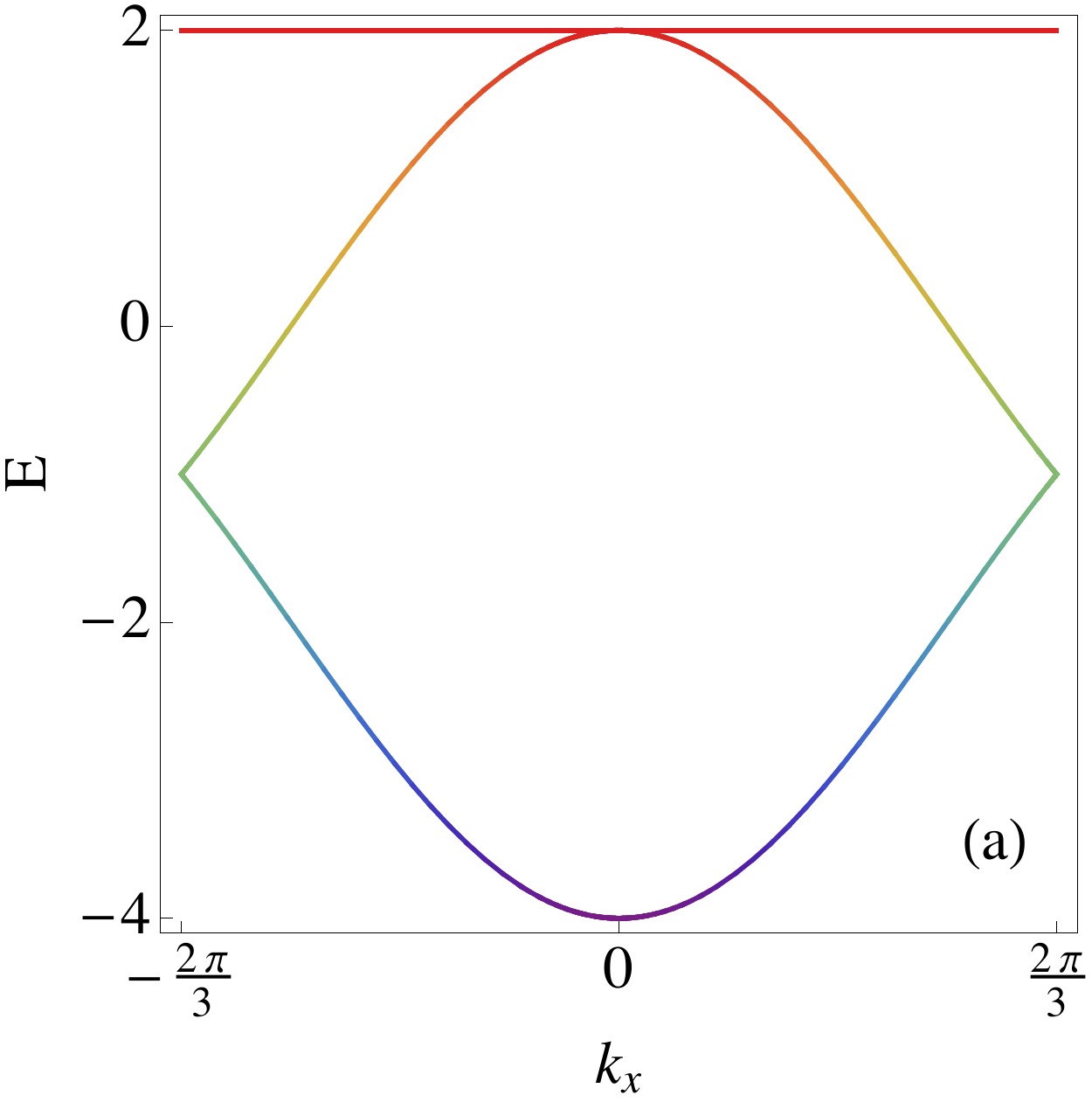}  %
\includegraphics[width=4cm]{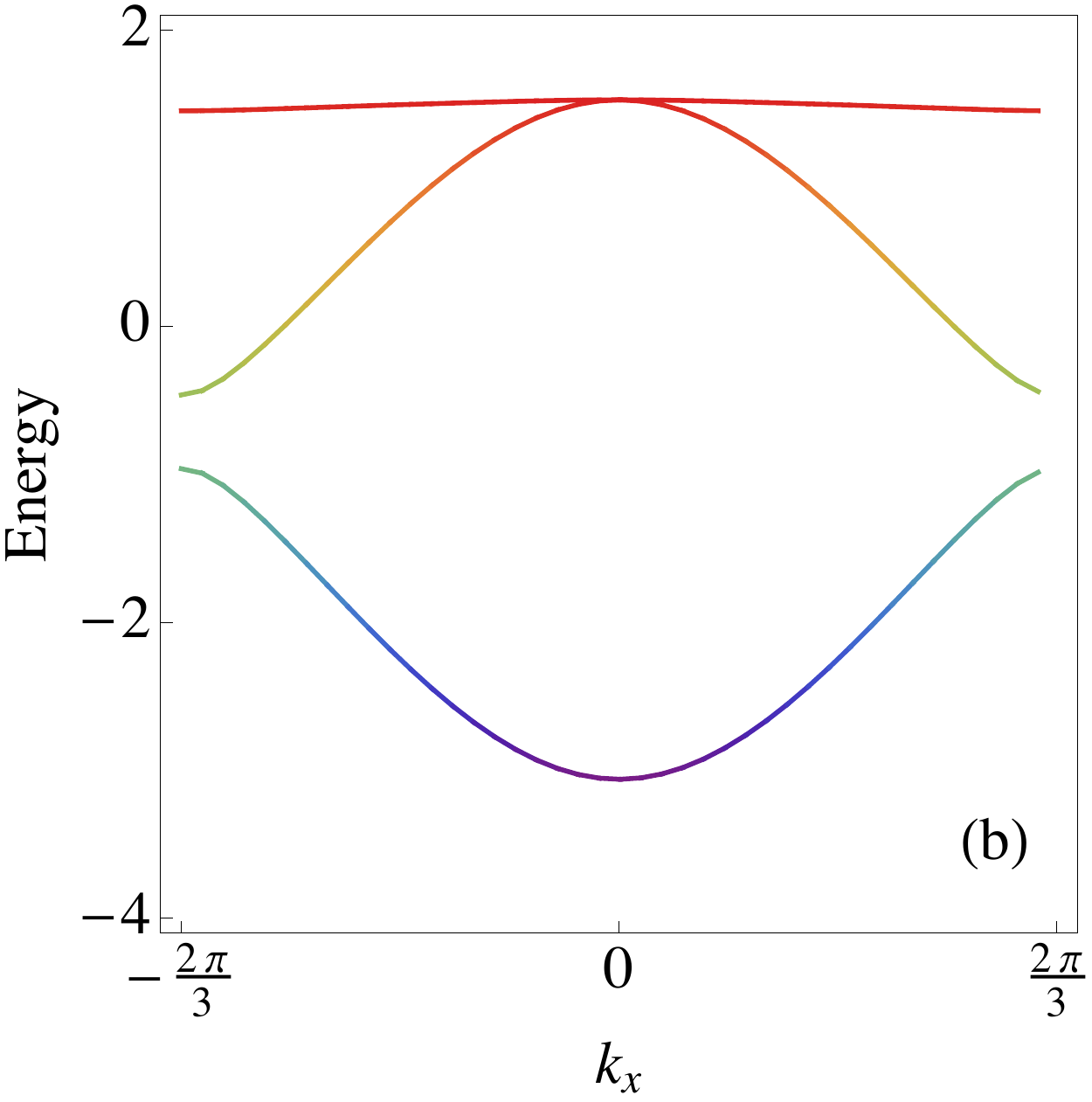}  %
\includegraphics[width=4cm]{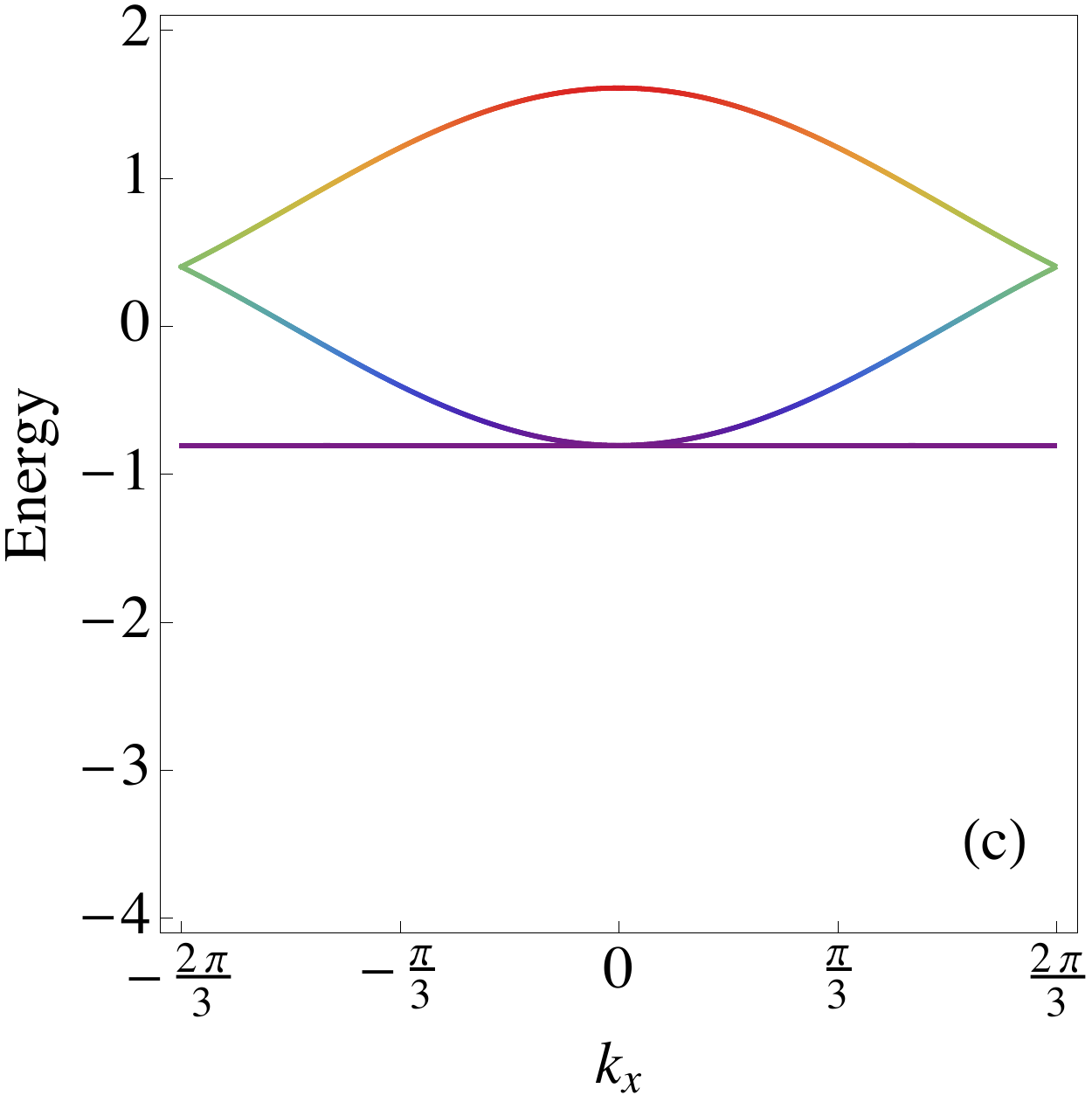}  %
\includegraphics[width=4cm]{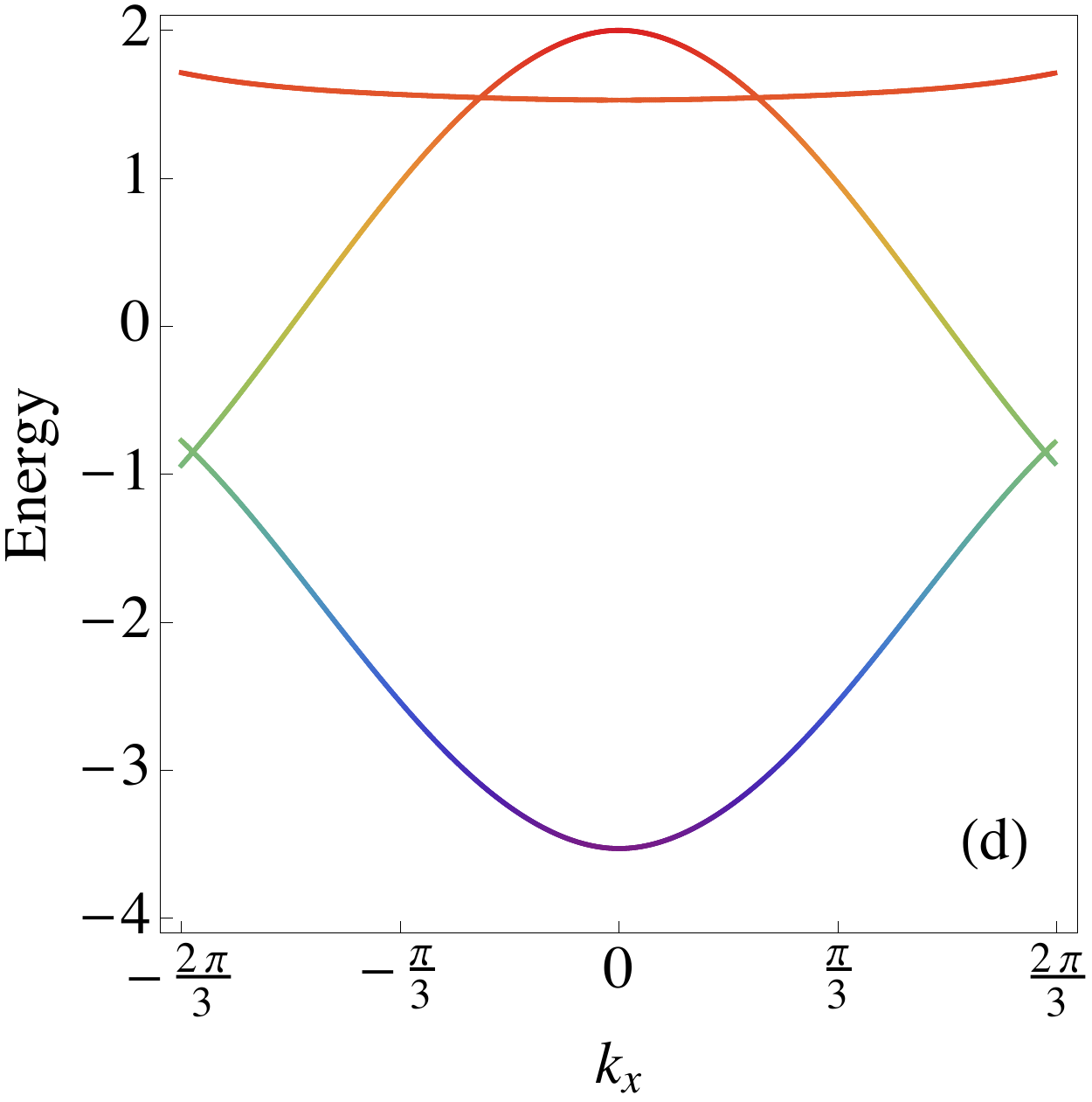}  %
\end{center}
\caption{(Color online) The Floquet-Bloch band structure of the kagome lattice embedded in a normally incident polarized light. 
The frequency of pump light is $\Omega = 6\tsh$. The color of the lines are the same as in Fig.\ref{fig:floquet-band-3d}.
(a) the band structure in equilibrium at $k_y=0.0$
(b) The Floquet-Bloch band structure with circularly polarized light $A_0=1.0$ at $k_y=0.0$
(c) The Floquet-Bloch band structure with circularly polarized light $A_0=3.8$ at $k_y=0.0$
(d) The Floquet-Bloch band structure with linearly polarized light $A_0=1.0$ at $k_y=0.0$. 
}
\label{fig:floquet-band-2d}
\end{figure}

Diagonalizing Eq.~(\ref{eq:htbk}) gives the following band structure,
\begin{align}
    E^{1,2}_{\bfk} = -\tsh [1 \pm \sqrt{4M_{\bfk}-3}], \quad
    E^{3}_{\bfk} = 2\tsh,
\end{align}
with $M_{\mathbf{k}}= \cos^{2}(\bfka_1)+ \cos^{2}(\bfka_2)+ \cos^{2}(\bfka_3)$. Band 1 (lowest energy band) and 2 (medium energy band) touch 
at two inequivalent Dirac points 
${\bf K}=(\frac{2\pi}{3},0)$ and ${\bf K'}(=-{\bf K})$ at the corners of the hexagonal Brillouin zone (BZ), band 2 (medium energy band) and 
3 (highest energy band) touch at the ${\bf \Gamma}$ point, 
resulting in a quadratic band crossing point. 
The band structure in $(k_x, k_y)$ space is shown in Fig.\ref{fig:floquet-band-3d}a, 
additionally, a cut along $k_y=0$ is shown in Fig.\ref{fig:floquet-band-2d}a.

\section{Periodic Drive under a laser field}
\label{sec:drive}
When the system is exposed to a normally incident laser field, the momentum is modified through the minimal coupling rule, $\mathbf{k} \rightarrow \mathbf{k} + \mathbf{A}(t)$, with $\mathbf{A}(t)$ the in-plane laser vector potential, and the Hamiltonian becomes time dependent,
%>>> time dependent tight-bing hamiltonian in momentum space
\begin{widetext}
    \begin{equation}\label{eq:htbk-t}
        {H}_{\mathbf{k}}(t) = -2 t_{\text h} \\
        \begin{pmatrix}
            0 & \cos[(\mathbf{k}+\mathbf{A}(t)) \cdot \mathbf{a}_{1}] & \cos[(\mathbf{k}+\mathbf{A}(t)) \cdot \mathbf{a}_{2}]\\
            \cos[(\mathbf{k}+\mathbf{A}(t)) \cdot \mathbf{a}_{1}] & 0 & \cos[(\mathbf{k}+\mathbf{A}(t)) \cdot \mathbf{a}_{3}]\\
            \cos[(\mathbf{k}+\mathbf{A}(t)) \cdot \mathbf{a}_{2}] & \cos[(\mathbf{k}+\mathbf{A}(t)) \cdot \mathbf{a}_{3}] & 0
        \end{pmatrix}.
    \end{equation}
\end{widetext}
In Eq.\eqref{eq:htbk-t}, we set Planck's constant $\hbar=1$, the speed of light $c=1$, and the charge of the electron, $e=1$, and adopt the Coulomb gauge by setting the scaler potential $\phi=0$. We ignore the tiny effect of magnetic field.  The units of energy are expressed in term of $t_h$, and we set $t_h=1$.

Throughout this paper, circularly polarized laser fields are expressed with the vector potential $\bfA(t)=A_0[\cos(\Omega t), -\sin(\Omega t)]$ and linear polarized laser fields are expressed with 
$\bfA(t) = A_0 [0, \cos(\Omega t)]$, where $A_0$  is the amplitude and $\Omega$ the frequency of the laser.

\section{Floquet Theory}
\label{sec:Floquet}

In this paper, we study monochromatic (single frequency) light, which renders the Hamiltonian time periodic $H(t) = H(t+T)$ where $T$ is the period of the drive. Hence, Floquet's theory can be used.\cite{Floquet:1883} The Floquet eigenfunction for the time periodic Hamiltonian can be expressed as,
\begin{equation}
     |\Psi_{k\alpha}(t)\rangle = e^{i\epsilon_{k\alpha} t} |\phi_{k\alpha}(t)\rangle,
\end{equation}
where $|\phi_{k\alpha}(t)\rangle = |\phi_{k\alpha}(t+T)\rangle$ are the Floquet quasimodes and 
$\epsilon_{k\alpha}$ is the corresponding quasienergy for band $\alpha$.
Substituting this form of the wavefunction into the time-dependent Schr\"odinger equation, 
%%% $i\frac{\partial}{\partial t}|\Psi_{k\alpha}(t)\rangle = H_{\bfk}(t) |\Psi_{k\alpha}(t)\rangle$
and defining the Floquet Hamiltonian operator as $\mathcal{H}(t) = H(t) - i{\partial}/{\partial t}$, one finds
\begin{equation}
     \mathcal{H}_{k}(t) |\phi_{k\alpha}(t)\rangle = \epsilon_{k\alpha} |\phi_{k\alpha}(t)\rangle.
\end{equation}
Here we restrict the quasienergy to be in the first Floquet zone, {\it i.e.}, $-\Omega/2 < \epsilon_{k\alpha} < \Omega/2$, and 
label the three bands using $\alpha=1, 2, 3$ by energy in ascending order. (Note that we have made use of a spin-independent coupling to the laser field so that all bands are 2-fold degenerate.  Henceforth, we suppress the spin degeneracy.)
Solving for the Floquet states in Fourier space,
\begin{equation}
     |\phi_{k\alpha}(t)\rangle = \sum_{m} e^{i m \Omega t} |\tilde{\phi}_{k\alpha}^m\rangle,
\end{equation}
where $m=0,\pm 1, \pm 2, \cdots $ and $|\tilde{\phi}_{k\alpha}^m\rangle$ is a three-component vector which obeys,
%>>> Floquet Hamiltonian in Fourier space
\begin{align}
     \sum_{m} (H_{nm} + m\Omega\delta_{nm}) |\tilde{\phi}_{k\alpha}^m\rangle = \epsilon_{k\alpha} |\tilde{\phi}_{k\alpha}^m\rangle,
\end{align}
with matrix elements of the Floquet Hamiltonian written as,
\begin{align}
    H_{nm}(k) &= \frac{1}{T}\int_{0}^{T} dt e^{-i(n-m)\Omega t} H(t) \nonumber\\
              &= \begin{pmatrix}
                  0 & H^{ab}_{nm}(k) & H^{ac}_{nm}(k) \\
                  H^{ab}_{nm}(k) & 0 & H^{bc}_{nm}(k) \\
                  H^{ac}_{nm}(k) & H^{bc}_{nm}(k) & 0
                 \end{pmatrix}.
\end{align}
Here $m$ and $n$ are integers ranging from $-\infty$ to $\infty$. Thus, the Floquet matrix is an infinite dimensional time-independent matrix.  In this paper, we consider the laser frequency to be comparable to or larger than the bandwidth of the system, so a
truncation of the components to be in $m,n=-1,0,1$ is a good approximation.  We have numerically verified that including a larger range of $m,n$ has a very small numerical impact on our results in Figs.\ref{fig:floquet-band-3d} and \ref{fig:floquet-band-2d}.

\subsection{Circularly Polarized Light}
For circularly polarized light with vector potential $\bfA(t)=A_0[\cos(\Omega t), -\sin(\Omega t)]$, the matrix elements of the Floquet-Bloch Hamiltonian are
\begin{align}
     H^{ab}_{nm} &= -t_{\text h} e^{i(m-n)\pi/2} \left[f(k_1, A_0) + f(-k_1, -A_0)\right],  \nonumber\\
     H^{ac}_{nm} &= -t_{\text h} e^{i(m-n)\pi/6} \left[f(k_2, A_0) + f(-k_2, -A_0)\right],  \nonumber\\
     H^{bc}_{nm} &= -t_{\text h} e^{i(n-m)\pi/6} \left[f(k_3, A_0) + f(-k_3, -A_0)\right],
\end{align}
where $f(k_i, x) = \mathcal{J}_{m-n}(x) e^{i \bfka_i}$ with $\mathcal{J}_m(x)$ the order-$m$ Bessel function of the first kind.
Diagonalizing the time independent Hamiltonian will give one the Floquet-Bloch band structure.
Figure \ref{fig:floquet-band-3d}(a-c) shows the band structure for circularly polarized light with amplitudes $A_0=0.0, 1.0, 3.8$ and drive frequency $\Omega=6$.  

The dominant features of the band structure can be understood by considering the effective Hamiltonian at large $\Omega$. 
Starting from $\mathcal{T}\exp[-i\int_0^T H(t) dt] = \exp[i H_{\mathrm{eff}} T]$ 
with $\mathcal{T}$ the time-ordering operator, and taking the high frequency limit,
\begin{equation}
     H_{\mathrm{eff}} = \frac{1}{T} \int_0^T dt H(t) + \mathcal{O}(1/\Omega).
\end{equation}
At leading order in $1/\Omega$,  $H_{\mathrm{eff}} = \mathcal{J}_0 (A_0) H_0$, which means that the driven band structure is renormalized by a scale factor of the zero-$th$ order Bessel function, $\mathcal{J}_0(A_0)$. 
For small amplitude, the band structure is scaled by $\mathcal{J}_0(1.0) = 0.7652$ as shown in the 
Fig.\ref{fig:floquet-band-3d}b and Fig.\ref{fig:floquet-band-2d}b.
Increasing the amplitude will lead to a negative value of the Bessel function $\mathcal{J}_0(3.8)=-0.4026$, 
where the bands are scaled $|\mathcal{J}_0(3.8)|$ and inverted as shown in Fig.\ref{fig:floquet-band-3d}c and Fig.\ref{fig:floquet-band-2d}c.  If one desires a ``low-energy" band that is flat for enhanced interaction effects, choosing the amplitude $A_0$ optimally may help. (Of course, one will need to be careful with controlling the occupation of the bands, which may require ``reservoir engineering".\cite{Iadecola_2:prb15,Iadecola:prb15,Seetharam:prx15})

We note that the gap opening at the ${\bf K}({\bf K'})$ points can not be captured by expanding the effective Hamiltonian to zeroth order in $1/\Omega$ because the zeroth order terms only constitute a rescaling of the bands, and therefore the time-reversal 
symmetry is preserved. A further expansion to order $1/\Omega$ (to include the one-photon processes) will break the time reversal symmetry and 
open a gap.\cite{Haldane-prl61-1988,sentef2015theory}  As we will see in the following sections, the gapped bands will acquire a finite Chern number and will leave a signature in the finite frequency transverse optical conductivity, $\sigma_{xy}(\omega)$.

%For linear polarized with vector potential $\bfA(t) = A_0 cos(\Omega t) (cos\theta, sin\theta)$
%with $\theta$ is the polarization direction of the pump light. Hereafter $\theta = \pi/2$ is used through out the paper unless otherwise mentioned.
%The vector potential of linear polarized light is set to be $\bfA(t) = A_0 [0, cos(\omega t)]$.
%\begin{align}
%     H^{ab}_{nm} &= -t_{\text h} e^{i(m-n)\pi/2} \left[f(k_1, A_0  ) + f(-k_1, -A_0  )\right] \nonumber\\
%     H^{ac}_{nm} &= -t_{\text h} e^{i(m-n)\pi/2} \left[f(k_2, \frac{A_0}{2}) + f(-k_2, -\frac{A_0}{2})\right] \nonumber\\
%     H^{bc}_{nm} &= -t_{\text h} e^{i(n-m)\pi/2} \left[f(k_3, \frac{A_0}{2}) + f(-k_3, -\frac{A_0}{2})\right]
%\end{align}
\subsection{Linearly Polarized Light}

For linearly polarized light with vector potential $\bfA(t) = A_0 [0, \cos(\Omega t)]$ the matrix elements of the Floquet-Bloch Hamiltonian are,
\begin{align}
     H^{ab}_{nm} &= -t_{\text h} i^{m-n} \left[\delta_{n,m} 2 \cos(\bfk \cdot \bfa_1) \right],\nonumber\\
     H^{ac}_{nm} &= -t_{\text h} i^{m-n} \left[f(k_2, \sqrt{3} A_0/2) + f(-k_2, - \sqrt{3} A_0/2)\right], \nonumber\\
     H^{bc}_{nm} &= -t_{\text h} i^{m-n} \left[f(k_3, \sqrt{3} A_0/2) + f(-k_3, - \sqrt{3} A_0/2)\right],
\end{align}
where $f(k_i,x)= \mathcal{J}_{m-n}(x) e^{i \bfka_i}$ with $\mathcal{J}_m(x)$ the order-$m$ Bessel function of the first kind, as before. The band structure for linearly polarized light is shown in 
Fig.\ref{fig:floquet-band-3d}d and Fig.\ref{fig:floquet-band-2d}d.
The Dirac points at $1/3$ filling undergo a small shift and the quadratic band crossing point at $2/3$ filling splits into two Dirac points along the $x$ direction, perpendicular to the direction of the electric field of the light.
These dominant band features can again be explained by taking the high frequency limit, and analyzing how the bands are renormalized by the Bessel function. Finer structure results from mixing between different ``copies" (indexed by photon number, $m$ or $n$) of the Floquet-Bloch states. 

We can also consider the case of linearly polarized light along a general direction in the $x$-$y$ plane with vector potential $\bfA(t) = A_0 \cos(\Omega t) (\cos\theta, \sin\theta)$.  We find that:
\begin{enumerate}
\item For $\theta=\pi/2$ (and symmetry-related directions), the Dirac points will make a tiny shift and remain gapless, as discussed above. 
%     Here we need to keep the quadratic terms in the low energy Hamiltonian around $\bfk$($\bfk'$) point.
\item For $\theta\ne \pi/2$ (and symmetry-related directions), the lower two bands will open a gap.
\item The quadratic band crossing point is split into two Dirac points perpendicular to the polarization direction of the pump light.
\end{enumerate}

Moreover, since the Hamiltonian matrix elements are real numbers, the finite-frequency transverse optical Hall conductance, $\sigma_{xy}(\omega)$, must also be zero (see next section), generalizing the well-known result for the (zero-frequency) Hall conductance that guarantees it is zero if time-reversal symmetry is not broken.  We also verified this numerically.  The band structure of the kagome lattice model with linear polarized light (for $\theta=\pi/2$) is shown in Fig.\ref{fig:floquet-band-3d}d and Fig.\ref{fig:floquet-band-2d}d.

\section{LOW ENERGY EFFECTIVE HAMILTONIAN FOR quadratic band crossing point}
\label{sec:low_energy}
% 1. k dot p Hamiltonian around QTBC points K=(0, 0)  
%     (1) quadratic touch hamiltonian is repreduced as d_i = d_z = -1/2. d_x = 1/2
%     (2) the magus expansion Hamiltonian
%         plot the figs to show that the low energy Hamiltonian can reproduce the behavior of exact ones
%     (3) effect of linear polarization
%         linear polarization will open split a quadratic touching points into two Dirac points
%         (3.1) gap size dependence of polarization direction \theta and A0
%         (3.2) Dirac point position dependence of theta and A0
%     (4) effect of circular polarization
%         (4.1) circularly polarized light will shift the bands and the qbcp is not open.
%     (5) the final conclusion is "1/Omega" terms is not included in the energy spectra, only second order terms comes in.
%%%%%%%%%%%%%%%%%%%%%
% TODO: how to show that the berry curvature for quadratic touch is zero 
%       the time reversal symmetry broken
% ANS : The quadratic band crossing point will open a gap of around A^4 / (8 Omega)
%%%%%%%%%%%%%%%%%%%%%

From the previous section, we see that the dominant band structure properties of the Floquet-Bloch bands can be understood by the time averaged Hamiltonian.  However, to see how the quadratic band touching point and Dirac band touching points respond to the circularly and linearly polarized light, it is helpful to derive an effective low-energy theory.  Because the Dirac band touching point has been much discussed in the literature,\cite{} here we focus on the quadratic band touching point, which to the best of our knowledge has not been studied in detail before.

The effective Hamiltonian describing the quadratic band touching point (in the absence of the drive) is given by,
\begin{equation}\label{eq:QBP}
    H_{\qbc} = -\tsh
    \begin{pmatrix}
        k_x^2 - 2 & - k_x k_y \\
       -k_x k_y   & k_y^2 - 2
    \end{pmatrix}.
\end{equation}
Technical details related to the derivation of the effective Hamiltonian, Eq.\eqref{eq:QBP}, are given in Appendix \ref{app:low}. To second order in $k_x \ll 1, k_y \ll 1$, and $A_0 \ll 1$ for this specific model Hamiltonian, 
the effective Hamiltonian derived from the Floquet Hamiltonian is consistent with the one obtained by setting $\bfk \rightarrow \bfk + \bfA(t)$ in Eq.\eqref{eq:QBP}. (Though, this is not generally true.)  By comparing Eq.\eqref{eq:QBP} with the general quadratic band touching Hamiltonian,\cite{Sun:prl09}
\begin{equation}\label{eq:QBP_gen}
    H_{\qbc}(k) = d_I I + d_x \sigma_x + d_z \sigma_z,
\end{equation}
one can determine the coefficients, where $I$ is identity matrix, and $\sigma_x$ and $\sigma_z$ are two real Pauli matrices along $x$ and $z$, respectively, and
      $d_I =   t_I (k_x^2 + k_y^2)$, 
      $d_x = 2 t_x k_x k_y$, and
      $d_z =   t_z (k_x^2 - k_y^2)$.
By comparing Eq.\eqref{eq:QBP} and Eq.\eqref{eq:QBP_gen}, we find $t_I = t_z = -\tsh/2$, and $t_x = \tsh/2$, which demonstrates 
the preservation of $C_6$ rotational symmetry and the breaking of particle hole symmetry.\cite{Sun:prl09}

The time-dependent Hamiltonian $H_{\qbc}(k, t)$ is derived from Eq.\eqref{eq:QBP} by setting $\bfk \rightarrow \bfk + \bfA(t)$.
By expanding the Hamiltonian to first order in $1/\Omega$,
we arrive at the Floquet-Magus expansion,\cite{Polkovnikov:rmp11,D?Alessio201319}
\begin{align}\label{eq:Heff}
     H_{\mathrm{eff}} = H^0 + &\sum_n \frac{1}{n\Omega} ([H^n, H^{-n}] \nonumber\\
                                 &+ [H^{-n}-H^{n}, H^0]) + \mathcal{O}(1/\Omega^2),
\end{align}
where $H^n = \frac{1}{T} \int_0^T dt H_{\qbc}(t) e^{-i n \Omega t}$ is the $n$-th Fourier component, and can be thought of as a ``dressing" of the state by photons.

\subsection{Linearly Polarized Light}
Linear polarized light along a general direction in the plane can be expressed as $\bfA(t) = A_0 \cos(\Omega t) (\cos\theta , \sin\theta)$.  When incorporated into the Hamiltonian, the quadratic band crossing point will split into two Dirac points, as shown in Fig.\ref{fig:floquet-band-3d}(d) and Fig.\ref{fig:floquet-band-2d}(d).  The magnitude of the gap at the ${\bf \Gamma}$ point can be obtained from the low-energy form of the Hamiltonian, Eq.\eqref{eq:QBP}, and is $A_0^2/2$, independent of the frequency $\Omega$ and the polarization direction $\theta$. The Dirac points are situated at $\pm (-A_0 \sin\theta, A_0 \cos\theta)/\sqrt{2}$, which depends only on the amplitude and polarization direction of light.

One can obtain these results by taking the limit $\Omega \gg 6t_h$ and $A_0 \ll 1$, where the effective Hamiltonian is 
$H_{\mathrm{eff}} = \frac{1}{T} \int_0^T dt H_{\qbc}(t) = \bar{d}_I I + \bar{d}_x \sigma_x + \bar{d}_z \sigma_z$, with 
      $\bar{d}_I = \tsh (k_x^2 + k_y^2 + A_0^2/2) / 2$, 
      $\bar{d}_x = \tsh (2 k_x k_y + A_0^2 \sin(2\theta)/2)$,
      $\bar{d}_z = \tsh (k_x^2 - k_y^2 + A_0^2 \cos(2\theta)/2) / 2$.
Diagonalizing this $2 \times 2$ Hamiltonian will give us the two eigenenergies, 
$E_{\pm} = (-A_0^2 - 2k_x^2 - 2k_y^2 \pm \sqrt{[A_0^2 - 2(k_x^2 + k_y^2)]^2 + 8A_0^2(k_x \cos\theta + k_y \sin\theta)^2})/4$, which clearly shows the splitting behavior and the gap at the $\Gamma$ point. From symmetry considerations, linearly polarized light preserves the time reversal symmetry, while the $C_6$ rotational symmetry is broken down to $C_2$, which allows a gap to open at the ${\bf \Gamma}$ point.\cite{Sun:prl09}

One can contrast this behavior of the quadratic band touching point with what happens at the Dirac points.  Around the Dirac points at $\mathbf{K}$ ($\mathbf{K'}$), one can derive the effective Hamiltonian, keeping terms up to quadratic order in momentum (leading corrections to the pure Dirac dispersion).  We find that the response of Dirac point depends on the polarization direction of the pump light: For $\theta=\pi/2$ (and symmetry related directions), the Dirac points will undergo a small shift and remain gapless.
%shift by $\pm A_0/8\sqrt{3}$, $\mathbf{K} \rightarrow \mathbf{K}-A_0/8\sqrt{3}$  and $\mathbf{K'} \rightarrow \mathbf{K'}+A_0/8\sqrt{3}$ and remain gapless.  
Away from $\theta = \pi/2$ (and symmetry related directions), the lower two bands will open a gap.  These results were earlier highlighted near the end of Sec.~\ref{sec:Floquet}.

% Classical picture: oscillating dipoles in x diraction will generate magnetic field in pependicular axis y.

\subsection{Circularly Polarized Light}
The effect of circularly polarized light on the quadratic band touching point is rather different from the case of linearly polarized light. Taking the high frequency limit and performing the time average, the zeroth-order Hamiltonian $H_{\mathrm{eff}} = H^0 + A_0 I  /2$ shows that the band will shift by $A_0/2$ and no gap is opened.
Furthermore, if one incorporates the one-photon absorption and emission terms in the effective Hamiltonian, its correction at ${\bf \Gamma}$ vanishes.  Thus the leading correction to the band structure is of order $\mathcal{O}(1/\Omega^2)$.  Only through the two-photon  absorption and emission processes [$n$=2 in Eq.\eqref{eq:Heff}] will open a gap proportional to $A_0^4/\Omega$ at the quadratic band crossing point. By contrast, the gap at the Dirac points $\mathbf{K}$ and $\mathbf{K'}$ are proportional to $A_0^2/\Omega$.

\section{CHERN NUMBER AND OPTICAL HALL CONDUCTIVITY}
\label{sec:sigma}
% (1) main equation related to the optical Hall conductivity
% (2) main results for the ideal and quenched case
%     (2.1) show the main contribution is from \sigma_{12} while \sigma_{13} contribute tiny
%     (2.2) explain the peaks by showing the F_km as a function of "eup - edo - m Omega", 
%           while its peak will determine the peak if OHC
%     (2.3) show the berry curvature with 8 dirac cone.

%%%%%%%%%%%%%%%%%%%%%
% TODO: show the berry curvature behavior
%       show that the contribution is mainly come from m=0 and m=1 (graphene)
%       show the behavior of occupation
%
In this section, we would like to make contact with possible future experiments and compute an observable that would reveal some of the band features mentioned above.  Angle resolved photoemission spectroscopy (ARPES) is a natural candidate,\cite{Petersen:prl11,Wang:sci13,Mahmood:np16} but only occupied states can be detected with the method.  In the introduction, we have mentioned some of the inherent challenges in transport measurements on Floquet systems.  Here, we would like to address an alternative property, the finite-frequency transverse optical conductivity which is related to the Faraday rotation in an optical experiment,\cite{Aditiprb92-2015,Morimoto:prl09,O'Connell:prb82,Ikebe:prl10,Crassee:np11,Shimano:nc13} and whose zero frequency limit naturally reduces to the Chern number (in an equilibrium system). 

Within linear response theory, the optical conductivity is computed from the time averaged Berry curvature over one period,\cite{Aditiprb91-2015}
\begin{equation}
     \bar{F}_{k \alpha} = \frac{1}{T} \int_0^T 2 \Im [\langle\partial_y \phi_{k\alpha}(t)|\partial_x \phi_{k\alpha}(t)\rangle],
\end{equation}
where $\alpha$ is the band index and $ \Im $ denotes the imaginary part.  It is helpful to compute the Berry curvature in a gauge invariant form,\cite{Aditiprb91-2015} where
$\bar{F}_{k\alpha}$ is written as $\bar{F}_{k\alpha} = \sum_{\beta, m} F_{k}^{m,\alpha\beta}$ with
\begin{equation}
\label{eq:Fkm}
     F_k^{m,\alpha\beta} = i\left[A_{\beta x \alpha}^{-m} A_{\alpha y \beta}^m - A_{\beta y \alpha}^{-m} A_{\alpha x \beta}^m \right],
\end{equation}
where the Fourier transformed Berry ``vector potential" is\cite{Aditiprb91-2015}
\begin{equation}
     A_{\beta i \alpha}^m = \frac{1}{T} \int_0^T dt e^{-i m\Omega t} \langle\phi_{k\beta}(t)|\frac{\partial}{\partial k_i}\phi_{k\alpha}(t)\rangle.
\end{equation}
Details related to the numerical calculation of $F_k^m$ are given in the appendix \ref{app:byfkm}.

For a three-band model, the optical Hall conductance can be written as a sum of three terms describing the transitions among the three bands, generalizing the two-band results of Ref.[\onlinecite{Aditiprb91-2015, Aditiprb92-2015}],
\begin{align}
     \sigma_{xy}(\omega) &= \sigma_{xy}^{12}(\omega) + \sigma_{xy}^{13}(\omega) + \sigma_{xy}^{23}(\omega) \nonumber\\
                         &= \sum_{m=int} \left[\sigma_{xy}^{m,12}(\omega) + \sigma_{xy}^{m,13}(\omega) + \sigma_{xy}^{m,23}(\omega)\right],
\end{align}
with
\begin{align}
\label{eq:hall-m}
     \sigma_{xy}^{m,\alpha\beta}(\omega) &= -\frac{e^2}{2\pi h} \int d^2 k E_{m,\alpha\beta}^2 F_k^{m,\alpha\beta} \nonumber\\
                 &\times \frac{\omega^2 - E_{m, \alpha\beta}^2 - 2i\omega\delta}
                 {[\omega^2 - E_{m,\alpha\beta}^2]^2 + 4\omega^2\delta^2}[\rho_{k\alpha} - \rho_{k\beta}],
\end{align}
where
\begin{equation}
     E_{m,\alpha\beta} = E_{k\beta} - E_{k\alpha} - m\Omega.
\end{equation}

\begin{figure}[t]
\includegraphics[width=1.0\linewidth]{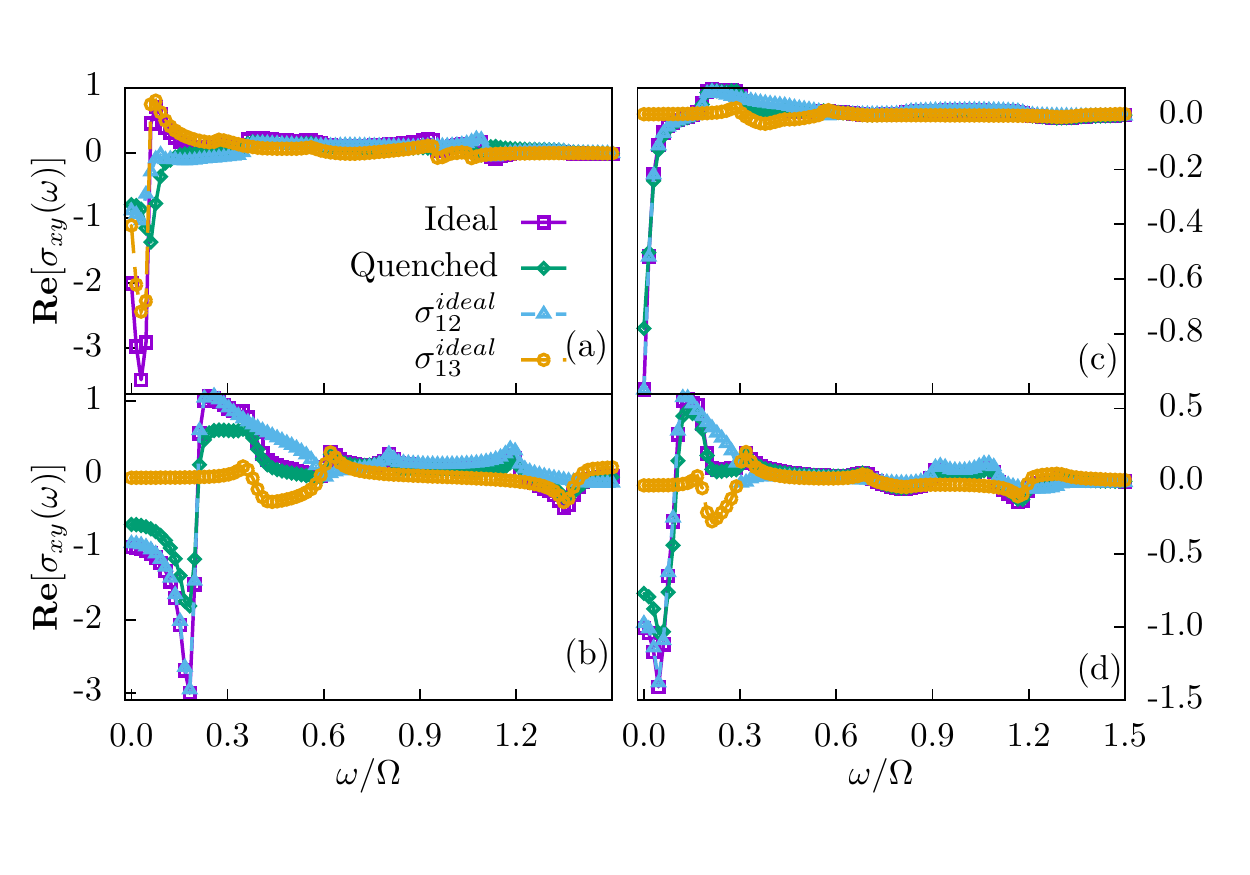}  %
\vspace{-12 mm}
\caption{(Color online) The optical Hall conductivity $\mathbf{Re}[\sigma_{xy}(\omega)]$ as a function of the frequency of the probe light, 
in units of $e^2/h$.
The driving laser frequency, laser amplitude and the Chern number (zero frequency limit of ${\sigma_{xy}(0) \over e^2/h}$) are 
(a) $\Omega=5, A_0=0.5, C=-2$, (b) $\Omega=5, A_0 = 1.5, C=-1$, (c) $\Omega=10, A_0 = 0.5, C=-1$, (d) $\Omega=10, A_0 = 1.5, C=-1$. 
A disorder broading $\delta=\Omega/100$ is chosen. For a description of ideal and quench cases, see text.}
\label{fig:opticalhall}
\end{figure}

\begin{figure}[t]
\includegraphics[width=1.0\linewidth]{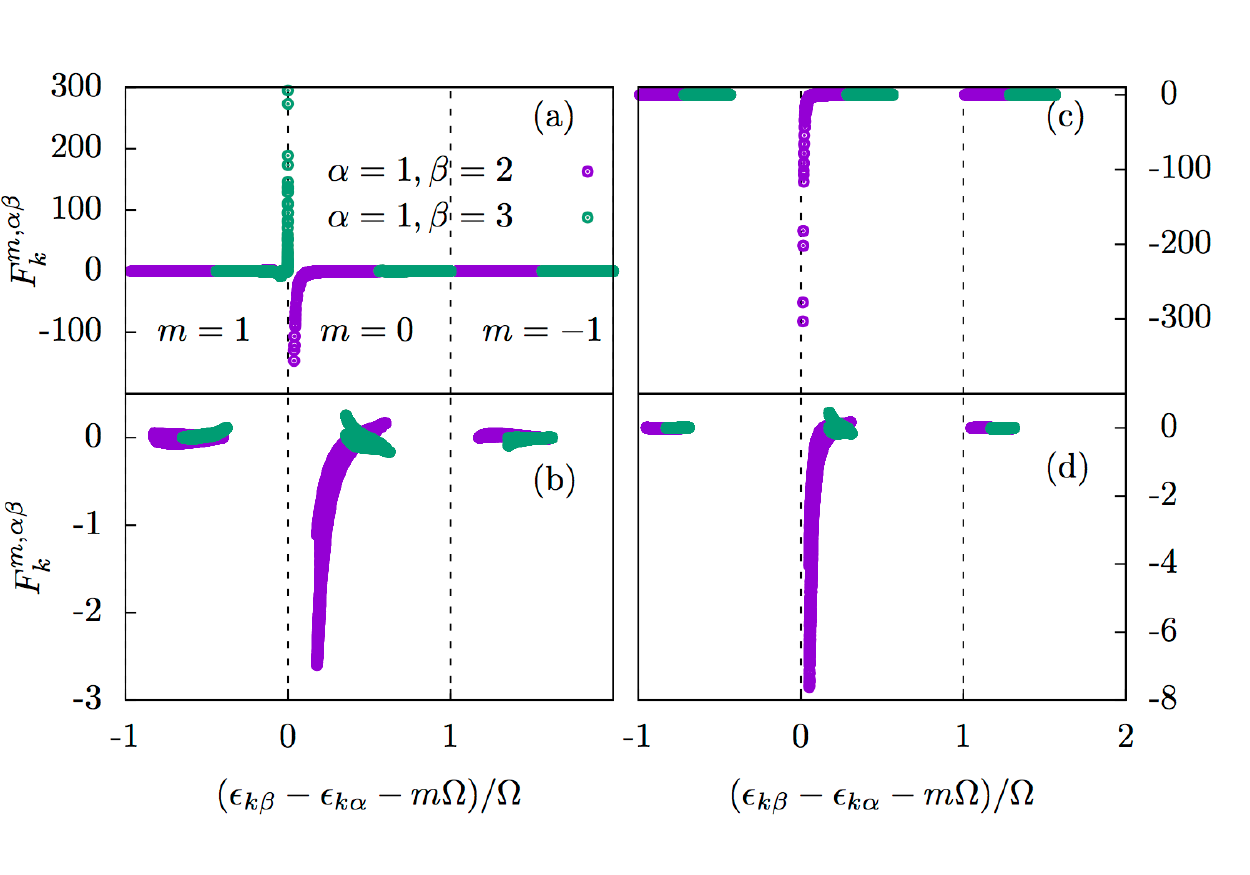}  %
\vspace{-10 mm}
\caption{(Color online) $F_k^{m, \alpha\beta}$ as a function of $(\epsilon_{k\beta}-\epsilon_{k\alpha} - m\Omega)/\Omega$ 
where band index $(\alpha,\beta)=(1,2), (1,3)$ at 
(a) $\Omega=5, A_0=0.5, C=-2$, (b) $\Omega=5, A_0 = 1.5, C=-1$, (c) $\Omega=10, A_0 = 0.5, C=-1$, (d) $\Omega=10, A_0 = 1.5, C=-1$. 
}
\label{fig:byfkm}
\end{figure}

The terms $\rho_{k\alpha} - \rho_{k\beta}$ in Eq.\eqref{eq:hall-m} describe the relative occupations of bands $\alpha$ and $\beta$ at wave vector $k$.  In our calculations, we assume that the system is initially (at zero time) in the ground state of the kagome lattice at $1/3$ filling and time evolve the system according to $H(t)$, which is the Hamiltonian Eq.\eqref{eq:htbk} modified by the vector potential $\bfA(t)$ [see Eq.\eqref{eq:htbk-t}] that drives the system into a Floquet-Bloch state.

We refer to the ``ideal" case as when the lowest Floquet-Bloch band of the system is fully occupied while the other two bands are empty (for all time).\cite{Aditiprb92-2015} For the quenched case, the occupation is defined as the overlap between the initial state ($t<0$) and the Floquet quasimode,
\begin{equation}
     \rho_{k\alpha}^{\text{quench}} = |\langle\phi_{k\alpha}(0)|\Psi_{in,k}\rangle|^2,
\end{equation}
where $|\Psi_{in,k}\rangle$ is the ground state wave function of the kagome lattice at 1/3 filling  before the laser is switched on at time zero.

Fig.\ref{fig:opticalhall} shows the optical Hall conductivity as a function of the frequency of {\em probe} light.  In the low frequency limit,
the dc Hall conductivity is proportional to the Chern number, $C$: $\sigma_{xy}^{ideal}(\omega=0) = C e^2 / h$ [see Eq.\eqref{eq:hall-m}], and is determined by the $\sum_{m}F_{k}^m$ in Eq.\eqref{eq:Fkm}. Here we used 7 copies $m=-3,\cdots,3$ in the Floquet Hamiltonian for calculations of optical Hall Conductivity. 
Away from the low frequency limit, the individual $F_k^m$ control the behavior.  One can see from Eq.\eqref{eq:hall-m} that the peaks in the integrand are around $\omega \approx |\epsilon_{k\beta} - \epsilon_{k\alpha} - m\Omega|$, while the dominant $k$ are determined by the peaks in $F_k^m$.
The high frequency limit behavior is $\sigma_{xy}(\omega) \propto 1/\omega^2$.
To better see these contributions we plot the $F_k^m$ as a function of $(|\epsilon_{k\beta} - \epsilon_{k\alpha} - m\Omega|)/\Omega$ in Fig.\ref{fig:byfkm}. 
The main structure of the optical Hall conductance is determined by $F_k^{m=0,12}$, while $F_k^{m=0,13}$ contribute to the fine structure of it.  Thus, the $F_k^m$ decay quite rapidly as a function of $m$.  This implies that the dominant structure of the optical conductivity in Eq.\eqref{eq:hall-m} comes from $m=0$, and offers a strategy for easily computing the dominant features of $\sigma_{xy}(\omega)$.

\section{DISCUSSION AND CONCLUSIONS}
\label{sec:conclusions}

In this work, we theoretically study the Floquet-Bloch band structure of a three-band model that includes both Dirac and quadratic band touching points, as well as a flat band.  We investigated the effects of both circularly and linearly polarized light.  We focused on the response of the quadratic band touching point to the periodic drive.  The dominant effect of circularly polarized light is to open a small gap of order $A_0^4/\Omega$ from two-photon processes, while the dominant effect of the linearly polarized light is to split the quadratic band touching point into two Dirac points at a location of $(\pm A_0/\sqrt{2}, 0)$, where the quadratic band touching point was initially at the $\Gamma$ point.  The splitting is perpendicular to the direction of the electric field of the linearly polarized light. To better understand these results, we derived a low-energy effective theory that allows one to understand these features analytically.  In addition, we showed that the dominate modifications of the band structure, including the ``inversion" of the band order, can be understood from the large $\Omega$ limit, even when $\Omega$ is as small as the order of the bandwidth itself.

Finally, we computed the finite-frequency optical conductivity, $\sigma_{xy}(\omega)$, for various scenarios of circularly polarized light (there is no response from linearly polarized light due to the preservation of time-reversal symmetry), including an ``ideal case" where the lower of the three Floquet-Bloch bands is occupied and the higher two are empty, as well as a ``quench case" where the band occupation is computed numerically from an initial state in which the lowest band of the time-independent Hamiltonian is occupied, and the higher two are empty.  We analyzed the contributions in both cases to determine the dominant contributions from resonant transition between the various bands, including the Floquet-Bloch index, $m$.  We found that the weights from different $m$ decay very rapidly with $m$ which suggests a strategy for quickly estimating the optical conductivity for more general band structures with more bands.  Because of the challenges of probing topological Floquet-Bloch systems with electrical transport measurements, optical probes of solid-state systems offer an alternative.

While our results are based on a closed quantum system, if a bath (of phonons or magnetic excitations, for example) were present much of our results would be qualitatively unchanged.  If the bath is at finite temperature, for example, the results in Fig.\ref{fig:opticalhall} would be thermally smeared, as discussed in Refs.[\onlinecite{Aditiprb92-2015,Aditiprb91-2015}].  Eventually, the signal will decrease and smoothly wash out when the temperature is of order the bandwidth of the equilibrium system.  

\section*{Acknowledgements} 

 We acknowledge helpful discussions with H. Dehghani, A. Mitra, Qi Chen and Ming Xie, and we gratefully acknowledge funding from ARO grant W911NF-14-1-0579 and NSF DMR-1507621.
%\newpage

\begin{appendix}
\section{Derivation of the effective Hamiltonian near the ${\bf \Gamma}$ point}
\label{app:low}
Starting from the original time dependent Hamiltonian $H_{\bfk}(t)$ in Eq.\eqref{eq:htbk-t}, we use a three-steps procedure to obtain the effective Hamiltonian:
Writing the full Floquet Hamiltonian in Fourier space and truncating it to $m,n=-1,0,1$, will give us a $9\times9$ matrix for each $\bfk$-point. Next, take the eigenstates $V$ at the ${\bf \Gamma}$ point as the new basis states, and write the Floquet Hamiltonian in the new basis by the rotation 
$H_{n, m}^{(1)}(k) = V^\dagger \cdot H_{n,m}(k) \cdot V$, where $V$ is given by,
\begin{equation}
    V = \frac{1}{\sqrt{6}}
    \begin{pmatrix}
        \sqrt{2} &  \sqrt{3} & -1 \\
        \sqrt{2} & -\sqrt{3} & -1 \\
        \sqrt{2} & 0         &  2
    \end{pmatrix}.
\end{equation}
At $m=n$, the transformation above will rotate the leading order terms to the diagonal components of the matrix. The new Floquet Hamiltian can be expressed as,
\begin{equation}
    H^{(1)}(k) = 
    \begin{pmatrix}
        H_0^{(1)} -\mathds{1}\Omega& H_{-1}^{(1)} & 0 \\
        H_{+1}^{(1)} & H_0 & H_{-1}^{(1)} \\
        0 & H_{+1}^{(1)}   &  H_0^{(1)}+\mathds{1}\Omega
    \end{pmatrix}.
\end{equation}
where $\mathds{1}$ denotes a $3\times3$ identity matrix. We have used 
$H_{n,m}^{(1)}(k) = H_{n-m}^{(1)}(k)$, and dropped the two photon terms.
%=\frac{1}{T}\int_0^T H(t) exp[- i (n-m) \Omega t] dt

Then we expand each term in $H_{n, m}^{(1)}(k)$ to the combined second order of $k_x, k_y$ and $A_0$. Higher order terms are dropped. 
Taking the Hamiltonian incorporating circularized polarized light as an example,
\begin{equation}
    \frac{H_0^{(1)}(k)}{t_h} = 
    \begin{pmatrix}
        -4 + k_x^2 + k_y^2 + A_0^2   & k_x k_y / \sqrt{2} & \frac{-k_x^2 + k_y^2}{2\sqrt{2}} \\
        k_x k_y / \sqrt{2}           & 2 - k_x^2 - \frac{A_0^2}{2} & k_x k_y \\
        {(-k_x^2 + k_y^2)}/{(2\sqrt{2})} & k_x k_y         &  2 - k_y^2 - \frac{A_0^2}{2}
    \end{pmatrix},\nonumber
\end{equation}
\begin{align}
    \frac{H_{+1}^{(1)}(k)}{t_h} = &
    \begin{pmatrix}
        A_0(k_x + i k_y)   & \frac{A_0 (i k_x + k_y)}{2\sqrt{2}} & \frac{-A_0(k_x - i k_y)}{(2\sqrt{2})} \\
        \frac{ A_0(i k_x + k_y)}{(2\sqrt{2})} & - A_0 k_x & \frac{A_0 (i k_x + k_y)}{2} \\
        \frac{-A_0(k_x - i k_y)}{(2\sqrt{2})} & \frac{A_0 (i k_x + k_y)}{2}                &  -i A_0 k_y
    \end{pmatrix}\nonumber\\
    = & \frac{H_{-1}^{(1)\dagger}(k)}{t_h}.\nonumber
\end{align}

The second step is to downfold from a $9\times9$ to a $3\times3$ Hamiltonian matrix,
\begin{align}
    H_{\mathrm{eff}}(k) = & H_0^{(1)}(k) +  H_{-1}^{(1)}(k)\frac{1}{\epsilon - (H_0^{(1)}(k) + \mathds{1} \Omega)} H_{+1}^{(1)}(k) \nonumber\\
                          + & H_{+1}^{(1)}(k)\frac{1}{\epsilon - (H_0^{(1)}(k) - \mathds{1} \Omega)} H_{-1}^{(1)}(k).
\end{align}
We set $\epsilon \approx \epsilon_0 = 2 t_h$, which is the quadratic touching point energy at equilibrium.
% Here the Downfolding technique is consistent with the spirit to derive the effective Hamiltonian introduced by Kitagawa and coworkers.\cite{kitagawa}

The third and final step is to use the downfolding trick again to derive a $2\times2$ Hamiltonian from the $3\times3$ matrix in the previous step,
\begin{equation}
    H_{\mathrm{eff}}(k) = t_h 
    \begin{pmatrix}
         2 - k_x^2 - \frac{A_0^2}{2} & k_x k_y \\
          k_x k_y  &  2 - k_y^2 - \frac{A_0^2}{2}.
    \end{pmatrix}.
\end{equation}

\section{The evaluation of $F_k^m$}
\label{app:byfkm}
From Eq.\eqref{eq:Fkm},
\begin{align}
     F_k^{m,\alpha\beta} 
  &= i\left[A_{\beta x \alpha}^{-m} A_{\alpha y \beta}^m - A_{\beta y \alpha}^{-m} A_{\alpha x \beta}^m \right] \nonumber\\
  &= -2 \Im \left(\sum_{l}\langle\tilde{\phi}_{k\beta}^{l}|\partial_{k_x}\tilde{\phi}_{k\alpha}^{l-m}\rangle
                  \sum_{n}\langle\tilde{\phi}_{k\alpha}^{n}|\partial_{k_y}\tilde{\phi}_{k\beta}^{n+m}\rangle\right).
\end{align}
The two summations can be done separately using,
\begin{align} \label{fkm}
     \sum_{l} \langle\tilde{\phi}_{k\beta}^{l}|\partial_{k_x}|\tilde{\phi}_{k\alpha}^{l-m}\rangle &=
     \sum_{nl}\langle\tilde{\phi}_{k\beta}^{n}|(\partial_{k_x} h_k^{n-l-m})|\tilde{\phi}_{k\alpha}^{l}\rangle
             / (-E_{\alpha\beta}^{m}),\nonumber\\
     \sum_{n} \langle\tilde{\phi}_{k\alpha}^{n}|\partial_{k_y}|\tilde{\phi}_{k\beta}^{n+m}\rangle &=
     \sum_{nl}\langle\tilde{\phi}_{k\alpha}^{l}|(\partial_{k_y} h_k^{l+m-n})|\tilde{\phi}_{k\beta}^{n}\rangle
             / E_{\alpha\beta}^{m},
\end{align}
where $h_{k}^{m} = \frac{1}{T}\int_{0}^{T} dt e^{im\Omega t} h_{k}(t)$
and $E_{\alpha,\beta}^{m} = \epsilon_{k\beta} - (\epsilon_{k\alpha} + m\Omega)$.
In this way, $F_k^m$ is calculated gauge invariantly,
\begin{align}
    & \langle \phi_{k\beta}|\left[\nabla h_k(t)\right]|\phi_{k\alpha}\rangle \nonumber\\
  =\  & (i\partial_t+\epsilon_{k\alpha}-\epsilon_{k\beta})\langle\phi_{k\beta}|\nabla\phi_{k\alpha}\rangle 
    + [\nabla\epsilon_{k\alpha}]\delta_{\alpha\beta}\nonumber\\
  =\  & \sum_{n l}[(n-l)\Omega t + \epsilon_{k\alpha} - \epsilon_{k\beta}] \langle\tilde{\phi}^n_{k\beta}|\nabla\tilde{\phi}_{k\alpha}^l\rangle
      e^{-i(n-l)\Omega t}
\end{align}
where in the second line the Floquet equation 
$(i\partial_t - h_{k})|\phi_{k\alpha}(t)\rangle = \epsilon_{k\alpha}|\phi_{k\alpha}(t)\rangle$ is used, 
while in the third line we set $\alpha=\beta$ and Fourier expand 
the Floquet mode $|\phi_{k\alpha}(t)\rangle = \sum_{n}e^{i n\Omega t}|\tilde{\phi}_{k\alpha}^m\rangle$.
Multiplying $e^{i m \Omega t}$ in both sides and taking the integral over one period $T=1/\Omega$, we arrive at,
\begin{equation}
     \sum_{n} \langle\tilde{\phi}_{k\beta}^{n}|\nabla\tilde{\phi}_{k\alpha}^{n-m}\rangle =
     \sum_{nl}\langle\tilde{\phi}^n_{k\beta}|[\nabla h_k^{n-l-m}]\tilde{\phi}_{k\alpha}^{l}\rangle 
             / (-E_{\alpha\beta}^{m})
\end{equation}
which is the equation shown in Eq. (\ref{fkm}).
\end{appendix}

%\bibliography{kagomeref,disorder,Interface_DOE.bib,pyro111}{}

\begin{thebibliography}{92}%
\makeatletter
\providecommand \@ifxundefined [1]{%
 \@ifx{#1\undefined}
}%
\providecommand \@ifnum [1]{%
 \ifnum #1\expandafter \@firstoftwo
 \else \expandafter \@secondoftwo
 \fi
}%
\providecommand \@ifx [1]{%
 \ifx #1\expandafter \@firstoftwo
 \else \expandafter \@secondoftwo
 \fi
}%
\providecommand \natexlab [1]{#1}%
\providecommand \enquote  [1]{``#1''}%
\providecommand \bibnamefont  [1]{#1}%
\providecommand \bibfnamefont [1]{#1}%
\providecommand \citenamefont [1]{#1}%
\providecommand \href@noop [0]{\@secondoftwo}%
\providecommand \href [0]{\begingroup \@sanitize@url \@href}%
\providecommand \@href[1]{\@@startlink{#1}\@@href}%
\providecommand \@@href[1]{\endgroup#1\@@endlink}%
\providecommand \@sanitize@url [0]{\catcode `\\12\catcode `\$12\catcode
  `\&12\catcode `\#12\catcode `\^12\catcode `\_12\catcode `\%12\relax}%
\providecommand \@@startlink[1]{}%
\providecommand \@@endlink[0]{}%
\providecommand \url  [0]{\begingroup\@sanitize@url \@url }%
\providecommand \@url [1]{\endgroup\@href {#1}{\urlprefix }}%
\providecommand \urlprefix  [0]{URL }%
\providecommand \Eprint [0]{\href }%
\providecommand \doibase [0]{http://dx.doi.org/}%
\providecommand \selectlanguage [0]{\@gobble}%
\providecommand \bibinfo  [0]{\@secondoftwo}%
\providecommand \bibfield  [0]{\@secondoftwo}%
\providecommand \translation [1]{[#1]}%
\providecommand \BibitemOpen [0]{}%
\providecommand \bibitemStop [0]{}%
\providecommand \bibitemNoStop [0]{.\EOS\space}%
\providecommand \EOS [0]{\spacefactor3000\relax}%
\providecommand \BibitemShut  [1]{\csname bibitem#1\endcsname}%
\let\auto@bib@innerbib\@empty
%</preamble>
\bibitem [{\citenamefont {Moore}(2010)}]{Moore:nat10}%
  \BibitemOpen
  \bibfield  {author} {\bibinfo {author} {\bibfnamefont {J.~E.}\ \bibnamefont
  {Moore}},\ }\href@noop {} {\bibfield  {journal} {\bibinfo  {journal}
  {Nature}\ }\textbf {\bibinfo {volume} {464}},\ \bibinfo {pages} {194}
  (\bibinfo {year} {2010})}\BibitemShut {NoStop}%
\bibitem [{\citenamefont {Hasan}\ and\ \citenamefont
  {Kane}(2010)}]{Hasan:rmp10}%
  \BibitemOpen
  \bibfield  {author} {\bibinfo {author} {\bibfnamefont {M.~Z.}\ \bibnamefont
  {Hasan}}\ and\ \bibinfo {author} {\bibfnamefont {C.~L.}\ \bibnamefont
  {Kane}},\ }\href {\doibase 10.1103/RevModPhys.82.3045} {\bibfield  {journal}
  {\bibinfo  {journal} {Rev. Mod. Phys.}\ }\textbf {\bibinfo {volume} {82}},\
  \bibinfo {pages} {3045} (\bibinfo {year} {2010})}\BibitemShut {NoStop}%
\bibitem [{\citenamefont {Qi}\ and\ \citenamefont {Zhang}(2011)}]{Qi:rmp11}%
  \BibitemOpen
  \bibfield  {author} {\bibinfo {author} {\bibfnamefont {X.~L.}\ \bibnamefont
  {Qi}}\ and\ \bibinfo {author} {\bibfnamefont {S.~C.}\ \bibnamefont {Zhang}},\
  }\href@noop {} {\bibfield  {journal} {\bibinfo  {journal} {Rev. Mod. Phys.}\
  }\textbf {\bibinfo {volume} {83}},\ \bibinfo {pages} {1057} (\bibinfo {year}
  {2011})}\BibitemShut {NoStop}%
\bibitem [{\citenamefont {Ando}(2013)}]{Ando:jpsj13}%
  \BibitemOpen
  \bibfield  {author} {\bibinfo {author} {\bibfnamefont {Y.}~\bibnamefont
  {Ando}},\ }\href@noop {} {\bibfield  {journal} {\bibinfo  {journal} {J. Phys.
  Soc. Jpn.}\ }\textbf {\bibinfo {volume} {82}},\ \bibinfo {pages} {102001}
  (\bibinfo {year} {2013})}\BibitemShut {NoStop}%
\bibitem [{\citenamefont {Maciejko}\ and\ \citenamefont
  {Fiete}(2015)}]{Maciejko:np15}%
  \BibitemOpen
  \bibfield  {author} {\bibinfo {author} {\bibfnamefont {J.}~\bibnamefont
  {Maciejko}}\ and\ \bibinfo {author} {\bibfnamefont {G.~A.}\ \bibnamefont
  {Fiete}},\ }\href@noop {} {\bibfield  {journal} {\bibinfo  {journal} {Nat.
  Phys.}\ }\textbf {\bibinfo {volume} {11}},\ \bibinfo {pages} {385} (\bibinfo
  {year} {2015})}\BibitemShut {NoStop}%
\bibitem [{\citenamefont {Stern}(2015)}]{Stern:arcmp15}%
  \BibitemOpen
  \bibfield  {author} {\bibinfo {author} {\bibfnamefont {A.}~\bibnamefont
  {Stern}},\ }\href@noop {} {\enquote {\bibinfo {title} {Fractional topological
  insulators -- a pedagogical review},}\ } (\bibinfo {year} {2015}),\ \bibinfo
  {note} {arXiv:1509.02698}\BibitemShut {NoStop}%
\bibitem [{\citenamefont {Witczak-Krempa}\ \emph {et~al.}(2014)\citenamefont
  {Witczak-Krempa}, \citenamefont {Chen}, \citenamefont {Kim},\ and\
  \citenamefont {Balents}}]{witczak-krempa2014}%
  \BibitemOpen
  \bibfield  {author} {\bibinfo {author} {\bibfnamefont {W.}~\bibnamefont
  {Witczak-Krempa}}, \bibinfo {author} {\bibfnamefont {G.}~\bibnamefont
  {Chen}}, \bibinfo {author} {\bibfnamefont {Y.~B.}\ \bibnamefont {Kim}}, \
  and\ \bibinfo {author} {\bibfnamefont {L.}~\bibnamefont {Balents}},\ }\href
  {\doibase 10.1146/annurev-conmatphys-020911-125138} {\bibfield  {journal}
  {\bibinfo  {journal} {Annu. Rev. Condens. Matter Phys.}\ }\textbf {\bibinfo
  {volume} {5}},\ \bibinfo {pages} {57} (\bibinfo {year} {2014})}\BibitemShut
  {NoStop}%
\bibitem [{\citenamefont {Mesaros}\ and\ \citenamefont
  {Ran}(2013)}]{mesaros2013}%
  \BibitemOpen
  \bibfield  {author} {\bibinfo {author} {\bibfnamefont {A.}~\bibnamefont
  {Mesaros}}\ and\ \bibinfo {author} {\bibfnamefont {Y.}~\bibnamefont {Ran}},\
  }\href {\doibase 10.1103/PhysRevB.87.155115} {\bibfield  {journal} {\bibinfo
  {journal} {Phys. Rev. B}\ }\textbf {\bibinfo {volume} {87}},\ \bibinfo
  {pages} {155115} (\bibinfo {year} {2013})}\BibitemShut {NoStop}%
\bibitem [{\citenamefont {Chen}\ \emph {et~al.}(2013)\citenamefont {Chen},
  \citenamefont {Gu}, \citenamefont {Liu},\ and\ \citenamefont
  {Wen}}]{Chen:prb13}%
  \BibitemOpen
  \bibfield  {author} {\bibinfo {author} {\bibfnamefont {X.}~\bibnamefont
  {Chen}}, \bibinfo {author} {\bibfnamefont {Z.-C.}\ \bibnamefont {Gu}},
  \bibinfo {author} {\bibfnamefont {Z.-X.}\ \bibnamefont {Liu}}, \ and\
  \bibinfo {author} {\bibfnamefont {X.-G.}\ \bibnamefont {Wen}},\ }\href
  {\doibase 10.1103/PhysRevB.87.155114} {\bibfield  {journal} {\bibinfo
  {journal} {Phys. Rev. B}\ }\textbf {\bibinfo {volume} {87}},\ \bibinfo
  {pages} {155114} (\bibinfo {year} {2013})}\BibitemShut {NoStop}%
\bibitem [{\citenamefont {Sun}\ \emph {et~al.}(2009)\citenamefont {Sun},
  \citenamefont {Yao}, \citenamefont {Fradkin},\ and\ \citenamefont
  {Kivelson}}]{Sun:prl09}%
  \BibitemOpen
  \bibfield  {author} {\bibinfo {author} {\bibfnamefont {K.}~\bibnamefont
  {Sun}}, \bibinfo {author} {\bibfnamefont {H.}~\bibnamefont {Yao}}, \bibinfo
  {author} {\bibfnamefont {E.}~\bibnamefont {Fradkin}}, \ and\ \bibinfo
  {author} {\bibfnamefont {S.~A.}\ \bibnamefont {Kivelson}},\ }\href {\doibase
  10.1103/PhysRevLett.103.046811} {\bibfield  {journal} {\bibinfo  {journal}
  {Phys. Rev. Lett.}\ }\textbf {\bibinfo {volume} {103}},\ \bibinfo {eid}
  {046811} (\bibinfo {year} {2009})}\BibitemShut {NoStop}%
\bibitem [{\citenamefont {Meng}\ \emph {et~al.}(2010)\citenamefont {Meng},
  \citenamefont {Lang}, \citenamefont {Wessel}, \citenamefont {Assaad},\ and\
  \citenamefont {Muramatsu}}]{Meng:nat10}%
  \BibitemOpen
  \bibfield  {author} {\bibinfo {author} {\bibfnamefont {Z.~Y.}\ \bibnamefont
  {Meng}}, \bibinfo {author} {\bibfnamefont {T.~C.}\ \bibnamefont {Lang}},
  \bibinfo {author} {\bibfnamefont {S.}~\bibnamefont {Wessel}}, \bibinfo
  {author} {\bibfnamefont {F.~F.}\ \bibnamefont {Assaad}}, \ and\ \bibinfo
  {author} {\bibfnamefont {A.}~\bibnamefont {Muramatsu}},\ }\href@noop {}
  {\bibfield  {journal} {\bibinfo  {journal} {Nature}\ }\textbf {\bibinfo
  {volume} {464}},\ \bibinfo {pages} {847} (\bibinfo {year}
  {2010})}\BibitemShut {NoStop}%
\bibitem [{\citenamefont {Hohenadler}\ \emph {et~al.}(2011)\citenamefont
  {Hohenadler}, \citenamefont {Lang},\ and\ \citenamefont
  {Assaad}}]{Hohenadler:prl11}%
  \BibitemOpen
  \bibfield  {author} {\bibinfo {author} {\bibfnamefont {M.}~\bibnamefont
  {Hohenadler}}, \bibinfo {author} {\bibfnamefont {T.~C.}\ \bibnamefont
  {Lang}}, \ and\ \bibinfo {author} {\bibfnamefont {F.~F.}\ \bibnamefont
  {Assaad}},\ }\href {\doibase 10.1103/PhysRevLett.106.100403} {\bibfield
  {journal} {\bibinfo  {journal} {Phys. Rev. Lett.}\ }\textbf {\bibinfo
  {volume} {106}},\ \bibinfo {pages} {100403} (\bibinfo {year}
  {2011})}\BibitemShut {NoStop}%
\bibitem [{\citenamefont {Yu}\ \emph {et~al.}(2011)\citenamefont {Yu},
  \citenamefont {Xie},\ and\ \citenamefont {Li}}]{Yu:prl11}%
  \BibitemOpen
  \bibfield  {author} {\bibinfo {author} {\bibfnamefont {S.-L.}\ \bibnamefont
  {Yu}}, \bibinfo {author} {\bibfnamefont {X.~C.}\ \bibnamefont {Xie}}, \ and\
  \bibinfo {author} {\bibfnamefont {J.-X.}\ \bibnamefont {Li}},\ }\href
  {\doibase 10.1103/PhysRevLett.107.010401} {\bibfield  {journal} {\bibinfo
  {journal} {Phys. Rev. Lett.}\ }\textbf {\bibinfo {volume} {107}},\ \bibinfo
  {pages} {010401} (\bibinfo {year} {2011})}\BibitemShut {NoStop}%
\bibitem [{\citenamefont {Castro~Neto}\ \emph {et~al.}(2009)\citenamefont
  {Castro~Neto}, \citenamefont {Guinea}, \citenamefont {Peres}, \citenamefont
  {Novoselov},\ and\ \citenamefont {Geim}}]{Neto:rmp09}%
  \BibitemOpen
  \bibfield  {author} {\bibinfo {author} {\bibfnamefont {A.~H.}\ \bibnamefont
  {Castro~Neto}}, \bibinfo {author} {\bibfnamefont {F.}~\bibnamefont {Guinea}},
  \bibinfo {author} {\bibfnamefont {N.~M.~R.}\ \bibnamefont {Peres}}, \bibinfo
  {author} {\bibfnamefont {K.~S.}\ \bibnamefont {Novoselov}}, \ and\ \bibinfo
  {author} {\bibfnamefont {A.~K.}\ \bibnamefont {Geim}},\ }\href {\doibase
  10.1103/RevModPhys.81.109} {\bibfield  {journal} {\bibinfo  {journal} {Rev.
  Mod. Phys.}\ }\textbf {\bibinfo {volume} {81}},\ \bibinfo {pages} {109}
  (\bibinfo {year} {2009})}\BibitemShut {NoStop}%
\bibitem [{\citenamefont {Wen}\ \emph {et~al.}(2010)\citenamefont {Wen},
  \citenamefont {R\"uegg}, \citenamefont {Wang},\ and\ \citenamefont
  {Fiete}}]{Wen:prb10}%
  \BibitemOpen
  \bibfield  {author} {\bibinfo {author} {\bibfnamefont {J.}~\bibnamefont
  {Wen}}, \bibinfo {author} {\bibfnamefont {A.}~\bibnamefont {R\"uegg}},
  \bibinfo {author} {\bibfnamefont {C.-C.~J.}\ \bibnamefont {Wang}}, \ and\
  \bibinfo {author} {\bibfnamefont {G.~A.}\ \bibnamefont {Fiete}},\ }\href
  {\doibase 10.1103/PhysRevB.82.075125} {\bibfield  {journal} {\bibinfo
  {journal} {Phys. Rev. B}\ }\textbf {\bibinfo {volume} {82}},\ \bibinfo
  {pages} {075125} (\bibinfo {year} {2010})}\BibitemShut {NoStop}%
\bibitem [{\citenamefont {Weeks}\ and\ \citenamefont
  {Franz}(2010)}]{Weeks:prb10a}%
  \BibitemOpen
  \bibfield  {author} {\bibinfo {author} {\bibfnamefont {C.}~\bibnamefont
  {Weeks}}\ and\ \bibinfo {author} {\bibfnamefont {M.}~\bibnamefont {Franz}},\
  }\href {\doibase 10.1103/PhysRevB.81.085105} {\bibfield  {journal} {\bibinfo
  {journal} {Phys. Rev. B}\ }\textbf {\bibinfo {volume} {81}},\ \bibinfo
  {pages} {085105} (\bibinfo {year} {2010})}\BibitemShut {NoStop}%
\bibitem [{\citenamefont {Zhang}\ \emph {et~al.}(2009)\citenamefont {Zhang},
  \citenamefont {Ran},\ and\ \citenamefont {Vishwanath}}]{Zhang:prb09}%
  \BibitemOpen
  \bibfield  {author} {\bibinfo {author} {\bibfnamefont {Y.}~\bibnamefont
  {Zhang}}, \bibinfo {author} {\bibfnamefont {Y.}~\bibnamefont {Ran}}, \ and\
  \bibinfo {author} {\bibfnamefont {A.}~\bibnamefont {Vishwanath}},\ }\href
  {\doibase 10.1103/PhysRevB.79.245331} {\bibfield  {journal} {\bibinfo
  {journal} {Phys. Rev. B}\ }\textbf {\bibinfo {volume} {79}},\ \bibinfo {eid}
  {245331} (\bibinfo {year} {2009})}\BibitemShut {NoStop}%
\bibitem [{\citenamefont {Raghu}\ \emph {et~al.}(2008)\citenamefont {Raghu},
  \citenamefont {Qi}, \citenamefont {Honerkamp},\ and\ \citenamefont
  {Zhang}}]{Raghu:prl08}%
  \BibitemOpen
  \bibfield  {author} {\bibinfo {author} {\bibfnamefont {S.}~\bibnamefont
  {Raghu}}, \bibinfo {author} {\bibfnamefont {X.-L.}\ \bibnamefont {Qi}},
  \bibinfo {author} {\bibfnamefont {C.}~\bibnamefont {Honerkamp}}, \ and\
  \bibinfo {author} {\bibfnamefont {S.-C.}\ \bibnamefont {Zhang}},\ }\href
  {\doibase 10.1103/PhysRevLett.100.156401} {\bibfield  {journal} {\bibinfo
  {journal} {Phys. Rev. Lett.}\ }\textbf {\bibinfo {volume} {100}},\ \bibinfo
  {eid} {156401} (\bibinfo {year} {2008})}\BibitemShut {NoStop}%
\bibitem [{\citenamefont {R\"uegg}\ and\ \citenamefont
  {Fiete}(2012)}]{Ruegg:prl12}%
  \BibitemOpen
  \bibfield  {author} {\bibinfo {author} {\bibfnamefont {A.}~\bibnamefont
  {R\"uegg}}\ and\ \bibinfo {author} {\bibfnamefont {G.~A.}\ \bibnamefont
  {Fiete}},\ }\href {\doibase 10.1103/PhysRevLett.108.046401} {\bibfield
  {journal} {\bibinfo  {journal} {Phys. Rev. Lett.}\ }\textbf {\bibinfo
  {volume} {108}},\ \bibinfo {pages} {046401} (\bibinfo {year}
  {2012})}\BibitemShut {NoStop}%
\bibitem [{\citenamefont {Yang}\ \emph {et~al.}(2011)\citenamefont {Yang},
  \citenamefont {Zhu}, \citenamefont {Xiao}, \citenamefont {Okamoto},
  \citenamefont {Wang},\ and\ \citenamefont {Ran}}]{Yang:prb11a}%
  \BibitemOpen
  \bibfield  {author} {\bibinfo {author} {\bibfnamefont {K.-Y.}\ \bibnamefont
  {Yang}}, \bibinfo {author} {\bibfnamefont {W.}~\bibnamefont {Zhu}}, \bibinfo
  {author} {\bibfnamefont {D.}~\bibnamefont {Xiao}}, \bibinfo {author}
  {\bibfnamefont {S.}~\bibnamefont {Okamoto}}, \bibinfo {author} {\bibfnamefont
  {Z.}~\bibnamefont {Wang}}, \ and\ \bibinfo {author} {\bibfnamefont
  {Y.}~\bibnamefont {Ran}},\ }\href {\doibase 10.1103/PhysRevB.84.201104}
  {\bibfield  {journal} {\bibinfo  {journal} {Phys. Rev. B}\ }\textbf {\bibinfo
  {volume} {84}},\ \bibinfo {pages} {201104} (\bibinfo {year}
  {2011})}\BibitemShut {NoStop}%
\bibitem [{\citenamefont {R\"uegg}\ and\ \citenamefont
  {Fiete}(2011)}]{Ruegg11_2}%
  \BibitemOpen
  \bibfield  {author} {\bibinfo {author} {\bibfnamefont {A.}~\bibnamefont
  {R\"uegg}}\ and\ \bibinfo {author} {\bibfnamefont {G.~A.}\ \bibnamefont
  {Fiete}},\ }\href@noop {} {\bibfield  {journal} {\bibinfo  {journal} {Phys.
  Rev. B}\ }\textbf {\bibinfo {volume} {84}},\ \bibinfo {pages} {201103}
  (\bibinfo {year} {2011})}\BibitemShut {NoStop}%
\bibitem [{\citenamefont {Wang}\ \emph {et~al.}(2012)\citenamefont {Wang},
  \citenamefont {Dai},\ and\ \citenamefont {Xie}}]{Wang:epl12}%
  \BibitemOpen
  \bibfield  {author} {\bibinfo {author} {\bibfnamefont {L.}~\bibnamefont
  {Wang}}, \bibinfo {author} {\bibfnamefont {X.}~\bibnamefont {Dai}}, \ and\
  \bibinfo {author} {\bibfnamefont {X.~C.}\ \bibnamefont {Xie}},\ }\href
  {\doibase 110.1209/0295-5075/98/57001} {\bibfield  {journal} {\bibinfo
  {journal} {Euro. Phys. Lett..}\ }\textbf {\bibinfo {volume} {98}},\ \bibinfo
  {pages} {57001} (\bibinfo {year} {2012})}\BibitemShut {NoStop}%
\bibitem [{\citenamefont {Yoshida}\ \emph {et~al.}(2013)\citenamefont
  {Yoshida}, \citenamefont {Peters}, \citenamefont {Fujimoto},\ and\
  \citenamefont {Kawakami}}]{Yoshida:prb13}%
  \BibitemOpen
  \bibfield  {author} {\bibinfo {author} {\bibfnamefont {T.}~\bibnamefont
  {Yoshida}}, \bibinfo {author} {\bibfnamefont {R.}~\bibnamefont {Peters}},
  \bibinfo {author} {\bibfnamefont {S.}~\bibnamefont {Fujimoto}}, \ and\
  \bibinfo {author} {\bibfnamefont {N.}~\bibnamefont {Kawakami}},\ }\href
  {\doibase 10.1103/PhysRevB.87.085134} {\bibfield  {journal} {\bibinfo
  {journal} {Phys. Rev. B}\ }\textbf {\bibinfo {volume} {87}},\ \bibinfo
  {pages} {085134} (\bibinfo {year} {2013})}\BibitemShut {NoStop}%
\bibitem [{\citenamefont {Go}\ \emph {et~al.}(2012)\citenamefont {Go},
  \citenamefont {Witczak-Krempa}, \citenamefont {Jeon}, \citenamefont {Park},\
  and\ \citenamefont {Kim}}]{Go:prl12}%
  \BibitemOpen
  \bibfield  {author} {\bibinfo {author} {\bibfnamefont {A.}~\bibnamefont
  {Go}}, \bibinfo {author} {\bibfnamefont {W.}~\bibnamefont {Witczak-Krempa}},
  \bibinfo {author} {\bibfnamefont {G.~S.}\ \bibnamefont {Jeon}}, \bibinfo
  {author} {\bibfnamefont {K.}~\bibnamefont {Park}}, \ and\ \bibinfo {author}
  {\bibfnamefont {Y.~B.}\ \bibnamefont {Kim}},\ }\href {\doibase
  10.1103/PhysRevLett.109.066401} {\bibfield  {journal} {\bibinfo  {journal}
  {Phys. Rev. Lett.}\ }\textbf {\bibinfo {volume} {109}},\ \bibinfo {pages}
  {066401} (\bibinfo {year} {2012})}\BibitemShut {NoStop}%
\bibitem [{\citenamefont {Maciejko}\ \emph {et~al.}(2014)\citenamefont
  {Maciejko}, \citenamefont {Chua},\ and\ \citenamefont
  {Fiete}}]{Maciejko:prl14}%
  \BibitemOpen
  \bibfield  {author} {\bibinfo {author} {\bibfnamefont {J.}~\bibnamefont
  {Maciejko}}, \bibinfo {author} {\bibfnamefont {V.}~\bibnamefont {Chua}}, \
  and\ \bibinfo {author} {\bibfnamefont {G.~A.}\ \bibnamefont {Fiete}},\ }\href
  {\doibase 10.1103/PhysRevLett.112.016404} {\bibfield  {journal} {\bibinfo
  {journal} {Phys. Rev. Lett.}\ }\textbf {\bibinfo {volume} {112}},\ \bibinfo
  {pages} {016404} (\bibinfo {year} {2014})}\BibitemShut {NoStop}%
\bibitem [{\citenamefont {Pesin}\ and\ \citenamefont
  {Balents}(2010)}]{Pesin:np10}%
  \BibitemOpen
  \bibfield  {author} {\bibinfo {author} {\bibfnamefont {D.}~\bibnamefont
  {Pesin}}\ and\ \bibinfo {author} {\bibfnamefont {L.}~\bibnamefont
  {Balents}},\ }\href@noop {} {\bibfield  {journal} {\bibinfo  {journal}
  {Nature Phys.}\ }\textbf {\bibinfo {volume} {6}},\ \bibinfo {pages} {376}
  (\bibinfo {year} {2010})}\BibitemShut {NoStop}%
\bibitem [{\citenamefont {Kargarian}\ and\ \citenamefont
  {Fiete}(2010)}]{Kargarian:prb10}%
  \BibitemOpen
  \bibfield  {author} {\bibinfo {author} {\bibfnamefont {M.}~\bibnamefont
  {Kargarian}}\ and\ \bibinfo {author} {\bibfnamefont {G.~A.}\ \bibnamefont
  {Fiete}},\ }\href {\doibase 10.1103/PhysRevB.82.085106} {\bibfield  {journal}
  {\bibinfo  {journal} {Phys. Rev. B}\ }\textbf {\bibinfo {volume} {82}},\
  \bibinfo {pages} {085106} (\bibinfo {year} {2010})}\BibitemShut {NoStop}%
\bibitem [{\citenamefont {Shitade}\ \emph {et~al.}(2009)\citenamefont
  {Shitade}, \citenamefont {Katsura}, \citenamefont {Kune\v{s}}, \citenamefont
  {Qi}, \citenamefont {Zhang},\ and\ \citenamefont {Nagaosa}}]{Shitade:prl09}%
  \BibitemOpen
  \bibfield  {author} {\bibinfo {author} {\bibfnamefont {A.}~\bibnamefont
  {Shitade}}, \bibinfo {author} {\bibfnamefont {H.}~\bibnamefont {Katsura}},
  \bibinfo {author} {\bibfnamefont {J.}~\bibnamefont {Kune\v{s}}}, \bibinfo
  {author} {\bibfnamefont {X.-L.}\ \bibnamefont {Qi}}, \bibinfo {author}
  {\bibfnamefont {S.-C.}\ \bibnamefont {Zhang}}, \ and\ \bibinfo {author}
  {\bibfnamefont {N.}~\bibnamefont {Nagaosa}},\ }\href {\doibase
  10.1103/PhysRevLett.102.256403} {\bibfield  {journal} {\bibinfo  {journal}
  {Phys. Rev. Lett.}\ }\textbf {\bibinfo {volume} {102}},\ \bibinfo {eid}
  {256403} (\bibinfo {year} {2009})}\BibitemShut {NoStop}%
\bibitem [{\citenamefont {Dzero}\ \emph {et~al.}(2010)\citenamefont {Dzero},
  \citenamefont {Sun}, \citenamefont {Galitski},\ and\ \citenamefont
  {Coleman}}]{Dzero:prl10}%
  \BibitemOpen
  \bibfield  {author} {\bibinfo {author} {\bibfnamefont {M.}~\bibnamefont
  {Dzero}}, \bibinfo {author} {\bibfnamefont {K.}~\bibnamefont {Sun}}, \bibinfo
  {author} {\bibfnamefont {V.}~\bibnamefont {Galitski}}, \ and\ \bibinfo
  {author} {\bibfnamefont {P.}~\bibnamefont {Coleman}},\ }\href {\doibase
  10.1103/PhysRevLett.104.106408} {\bibfield  {journal} {\bibinfo  {journal}
  {Phys. Rev. Lett.}\ }\textbf {\bibinfo {volume} {104}},\ \bibinfo {pages}
  {106408} (\bibinfo {year} {2010})}\BibitemShut {NoStop}%
\bibitem [{\citenamefont {Budich}\ \emph {et~al.}(2012)\citenamefont {Budich},
  \citenamefont {Thomale}, \citenamefont {Li}, \citenamefont {Laubach},\ and\
  \citenamefont {Zhang}}]{Budich:prb12}%
  \BibitemOpen
  \bibfield  {author} {\bibinfo {author} {\bibfnamefont {J.~C.}\ \bibnamefont
  {Budich}}, \bibinfo {author} {\bibfnamefont {R.}~\bibnamefont {Thomale}},
  \bibinfo {author} {\bibfnamefont {G.}~\bibnamefont {Li}}, \bibinfo {author}
  {\bibfnamefont {M.}~\bibnamefont {Laubach}}, \ and\ \bibinfo {author}
  {\bibfnamefont {S.-C.}\ \bibnamefont {Zhang}},\ }\href {\doibase
  10.1103/PhysRevB.86.201407} {\bibfield  {journal} {\bibinfo  {journal} {Phys.
  Rev. B}\ }\textbf {\bibinfo {volume} {86}},\ \bibinfo {pages} {201407}
  (\bibinfo {year} {2012})}\BibitemShut {NoStop}%
\bibitem [{\citenamefont {Budich}\ \emph {et~al.}(2013)\citenamefont {Budich},
  \citenamefont {Trauzettel},\ and\ \citenamefont
  {Sangiovanni}}]{Budich:prb13}%
  \BibitemOpen
  \bibfield  {author} {\bibinfo {author} {\bibfnamefont {J.~C.}\ \bibnamefont
  {Budich}}, \bibinfo {author} {\bibfnamefont {B.}~\bibnamefont {Trauzettel}},
  \ and\ \bibinfo {author} {\bibfnamefont {G.}~\bibnamefont {Sangiovanni}},\
  }\href {\doibase 10.1103/PhysRevB.87.235104} {\bibfield  {journal} {\bibinfo
  {journal} {Phys. Rev. B}\ }\textbf {\bibinfo {volume} {87}},\ \bibinfo
  {pages} {235104} (\bibinfo {year} {2013})}\BibitemShut {NoStop}%
\bibitem [{\citenamefont {Wang}\ \emph {et~al.}(2011)\citenamefont {Wang},
  \citenamefont {Gu}, \citenamefont {Gong},\ and\ \citenamefont
  {Sheng}}]{Wang:prl11}%
  \BibitemOpen
  \bibfield  {author} {\bibinfo {author} {\bibfnamefont {Y.-F.}\ \bibnamefont
  {Wang}}, \bibinfo {author} {\bibfnamefont {Z.-C.}\ \bibnamefont {Gu}},
  \bibinfo {author} {\bibfnamefont {C.-D.}\ \bibnamefont {Gong}}, \ and\
  \bibinfo {author} {\bibfnamefont {D.~N.}\ \bibnamefont {Sheng}},\ }\href@noop
  {} {\bibfield  {journal} {\bibinfo  {journal} {Phys. Rev. Lett.}\ }\textbf
  {\bibinfo {volume} {107}},\ \bibinfo {pages} {146803} (\bibinfo {year}
  {2011})}\BibitemShut {NoStop}%
\bibitem [{\citenamefont {Sheng}\ \emph {et~al.}(2011)\citenamefont {Sheng},
  \citenamefont {Gu}, \citenamefont {Sun},\ and\ \citenamefont
  {Sheng}}]{Sheng11}%
  \BibitemOpen
  \bibfield  {author} {\bibinfo {author} {\bibfnamefont {D.~N.}\ \bibnamefont
  {Sheng}}, \bibinfo {author} {\bibfnamefont {Z.-C.}\ \bibnamefont {Gu}},
  \bibinfo {author} {\bibfnamefont {K.}~\bibnamefont {Sun}}, \ and\ \bibinfo
  {author} {\bibfnamefont {L.}~\bibnamefont {Sheng}},\ }\href {\doibase
  10.1038/ncomms1380} {\bibfield  {journal} {\bibinfo  {journal} {Nature
  Communications}\ }\textbf {\bibinfo {volume} {2}},\ \bibinfo {pages} {389}
  (\bibinfo {year} {2011})}\BibitemShut {NoStop}%
\bibitem [{\citenamefont {Sun}\ \emph {et~al.}(2011)\citenamefont {Sun},
  \citenamefont {Gu}, \citenamefont {Katsura},\ and\ \citenamefont
  {Das~Sarma}}]{Sun:prl11}%
  \BibitemOpen
  \bibfield  {author} {\bibinfo {author} {\bibfnamefont {K.}~\bibnamefont
  {Sun}}, \bibinfo {author} {\bibfnamefont {Z.}~\bibnamefont {Gu}}, \bibinfo
  {author} {\bibfnamefont {H.}~\bibnamefont {Katsura}}, \ and\ \bibinfo
  {author} {\bibfnamefont {S.}~\bibnamefont {Das~Sarma}},\ }\href@noop {}
  {\bibfield  {journal} {\bibinfo  {journal} {Phys. Rev. Lett.}\ }\textbf
  {\bibinfo {volume} {106}},\ \bibinfo {pages} {236803} (\bibinfo {year}
  {2011})}\BibitemShut {NoStop}%
\bibitem [{\citenamefont {Neupert}\ \emph
  {et~al.}(2011{\natexlab{a}})\citenamefont {Neupert}, \citenamefont {Santos},
  \citenamefont {Chamon},\ and\ \citenamefont {Mudry}}]{Neupert:prl11}%
  \BibitemOpen
  \bibfield  {author} {\bibinfo {author} {\bibfnamefont {T.}~\bibnamefont
  {Neupert}}, \bibinfo {author} {\bibfnamefont {L.}~\bibnamefont {Santos}},
  \bibinfo {author} {\bibfnamefont {C.}~\bibnamefont {Chamon}}, \ and\ \bibinfo
  {author} {\bibfnamefont {C.}~\bibnamefont {Mudry}},\ }\href {\doibase
  10.1103/PhysRevLett.106.236804} {\bibfield  {journal} {\bibinfo  {journal}
  {Phys. Rev. Lett.}\ }\textbf {\bibinfo {volume} {106}},\ \bibinfo {pages}
  {236804} (\bibinfo {year} {2011}{\natexlab{a}})}\BibitemShut {NoStop}%
\bibitem [{\citenamefont {Tang}\ \emph {et~al.}(2011)\citenamefont {Tang},
  \citenamefont {Mei},\ and\ \citenamefont {Wen}}]{Tang:prl11}%
  \BibitemOpen
  \bibfield  {author} {\bibinfo {author} {\bibfnamefont {E.}~\bibnamefont
  {Tang}}, \bibinfo {author} {\bibfnamefont {J.-W.}\ \bibnamefont {Mei}}, \
  and\ \bibinfo {author} {\bibfnamefont {X.-G.}\ \bibnamefont {Wen}},\ }\href
  {\doibase 10.1103/PhysRevLett.106.236802} {\bibfield  {journal} {\bibinfo
  {journal} {Phys. Rev. Lett.}\ }\textbf {\bibinfo {volume} {106}},\ \bibinfo
  {pages} {236802} (\bibinfo {year} {2011})}\BibitemShut {NoStop}%
\bibitem [{\citenamefont {Wu}\ \emph {et~al.}(2012)\citenamefont {Wu},
  \citenamefont {Bernevig},\ and\ \citenamefont {Regnault}}]{Wu_CI:prb12}%
  \BibitemOpen
  \bibfield  {author} {\bibinfo {author} {\bibfnamefont {Y.-L.}\ \bibnamefont
  {Wu}}, \bibinfo {author} {\bibfnamefont {B.~A.}\ \bibnamefont {Bernevig}}, \
  and\ \bibinfo {author} {\bibfnamefont {N.}~\bibnamefont {Regnault}},\ }\href
  {\doibase 10.1103/PhysRevB.85.075116} {\bibfield  {journal} {\bibinfo
  {journal} {Phys. Rev. B}\ }\textbf {\bibinfo {volume} {85}},\ \bibinfo
  {pages} {075116} (\bibinfo {year} {2012})}\BibitemShut {NoStop}%
\bibitem [{\citenamefont {Liu}\ \emph {et~al.}(2013)\citenamefont {Liu},
  \citenamefont {Kovrizhin},\ and\ \citenamefont {Bergholtz}}]{Liu:prb13}%
  \BibitemOpen
  \bibfield  {author} {\bibinfo {author} {\bibfnamefont {Z.}~\bibnamefont
  {Liu}}, \bibinfo {author} {\bibfnamefont {D.~L.}\ \bibnamefont {Kovrizhin}},
  \ and\ \bibinfo {author} {\bibfnamefont {E.~J.}\ \bibnamefont {Bergholtz}},\
  }\href {\doibase 10.1103/PhysRevB.88.081106} {\bibfield  {journal} {\bibinfo
  {journal} {Phys. Rev. B}\ }\textbf {\bibinfo {volume} {88}},\ \bibinfo
  {pages} {081106} (\bibinfo {year} {2013})}\BibitemShut {NoStop}%
\bibitem [{\citenamefont {Parameswaran}\ \emph {et~al.}(2013)\citenamefont
  {Parameswaran}, \citenamefont {Roy},\ and\ \citenamefont
  {Sondhi}}]{Parameswaran2013816}%
  \BibitemOpen
  \bibfield  {author} {\bibinfo {author} {\bibfnamefont {S.~A.}\ \bibnamefont
  {Parameswaran}}, \bibinfo {author} {\bibfnamefont {R.}~\bibnamefont {Roy}}, \
  and\ \bibinfo {author} {\bibfnamefont {S.~L.}\ \bibnamefont {Sondhi}},\
  }\href {\doibase http://dx.doi.org/10.1016/j.crhy.2013.04.003} {\bibfield
  {journal} {\bibinfo  {journal} {Comptes Rendus Physique}\ }\textbf {\bibinfo
  {volume} {14}},\ \bibinfo {pages} {816 } (\bibinfo {year}
  {2013})}\BibitemShut {NoStop}%
\bibitem [{\citenamefont {Qi}(2011)}]{Qi11}%
  \BibitemOpen
  \bibfield  {author} {\bibinfo {author} {\bibfnamefont {X.-L.}\ \bibnamefont
  {Qi}},\ }\href {\doibase 10.1103/PhysRevLett.107.126803} {\bibfield
  {journal} {\bibinfo  {journal} {Phys. Rev. Lett.}\ }\textbf {\bibinfo
  {volume} {107}},\ \bibinfo {pages} {126803} (\bibinfo {year}
  {2011})}\BibitemShut {NoStop}%
\bibitem [{\citenamefont {Neupert}\ \emph
  {et~al.}(2011{\natexlab{b}})\citenamefont {Neupert}, \citenamefont {Santos},
  \citenamefont {Ryu}, \citenamefont {Chamon},\ and\ \citenamefont
  {Mudry}}]{Neupert:prb11}%
  \BibitemOpen
  \bibfield  {author} {\bibinfo {author} {\bibfnamefont {T.}~\bibnamefont
  {Neupert}}, \bibinfo {author} {\bibfnamefont {L.}~\bibnamefont {Santos}},
  \bibinfo {author} {\bibfnamefont {S.}~\bibnamefont {Ryu}}, \bibinfo {author}
  {\bibfnamefont {C.}~\bibnamefont {Chamon}}, \ and\ \bibinfo {author}
  {\bibfnamefont {C.}~\bibnamefont {Mudry}},\ }\href {\doibase
  10.1103/PhysRevB.84.165107} {\bibfield  {journal} {\bibinfo  {journal} {Phys.
  Rev. B}\ }\textbf {\bibinfo {volume} {84}},\ \bibinfo {pages} {165107}
  (\bibinfo {year} {2011}{\natexlab{b}})}\BibitemShut {NoStop}%
\bibitem [{\citenamefont {Regnault}\ and\ \citenamefont
  {Bernevig}(2011)}]{Regnault:prx11}%
  \BibitemOpen
  \bibfield  {author} {\bibinfo {author} {\bibfnamefont {N.}~\bibnamefont
  {Regnault}}\ and\ \bibinfo {author} {\bibfnamefont {B.~A.}\ \bibnamefont
  {Bernevig}},\ }\href {\doibase 10.1103/PhysRevX.1.021014} {\bibfield
  {journal} {\bibinfo  {journal} {Phys. Rev. X}\ }\textbf {\bibinfo {volume}
  {1}},\ \bibinfo {pages} {021014} (\bibinfo {year} {2011})}\BibitemShut
  {NoStop}%
\bibitem [{\citenamefont {Kourtis}\ \emph {et~al.}(2014)\citenamefont
  {Kourtis}, \citenamefont {Neupert}, \citenamefont {Chamon},\ and\
  \citenamefont {Mudry}}]{Kourtis:prl14}%
  \BibitemOpen
  \bibfield  {author} {\bibinfo {author} {\bibfnamefont {S.}~\bibnamefont
  {Kourtis}}, \bibinfo {author} {\bibfnamefont {T.}~\bibnamefont {Neupert}},
  \bibinfo {author} {\bibfnamefont {C.}~\bibnamefont {Chamon}}, \ and\ \bibinfo
  {author} {\bibfnamefont {C.}~\bibnamefont {Mudry}},\ }\href {\doibase
  10.1103/PhysRevLett.112.126806} {\bibfield  {journal} {\bibinfo  {journal}
  {Phys. Rev. Lett.}\ }\textbf {\bibinfo {volume} {112}},\ \bibinfo {pages}
  {126806} (\bibinfo {year} {2014})}\BibitemShut {NoStop}%
\bibitem [{\citenamefont {Lindner}\ \emph {et~al.}(2011)\citenamefont
  {Lindner}, \citenamefont {Refael},\ and\ \citenamefont
  {Galitski}}]{Lindner:np11}%
  \BibitemOpen
  \bibfield  {author} {\bibinfo {author} {\bibfnamefont {N.~H.}\ \bibnamefont
  {Lindner}}, \bibinfo {author} {\bibfnamefont {G.}~\bibnamefont {Refael}}, \
  and\ \bibinfo {author} {\bibfnamefont {V.}~\bibnamefont {Galitski}},\ }\href
  {\doibase 10.1038/nphys1926} {\bibfield  {journal} {\bibinfo  {journal} {Nat.
  Phys.}\ }\textbf {\bibinfo {volume} {7}},\ \bibinfo {pages} {490} (\bibinfo
  {year} {2011})}\BibitemShut {NoStop}%
\bibitem [{\citenamefont {Kitagawa}\ \emph {et~al.}(2010)\citenamefont
  {Kitagawa}, \citenamefont {Berg}, \citenamefont {Rudner},\ and\ \citenamefont
  {Demler}}]{Kitagawa:prb10}%
  \BibitemOpen
  \bibfield  {author} {\bibinfo {author} {\bibfnamefont {T.}~\bibnamefont
  {Kitagawa}}, \bibinfo {author} {\bibfnamefont {E.}~\bibnamefont {Berg}},
  \bibinfo {author} {\bibfnamefont {M.}~\bibnamefont {Rudner}}, \ and\ \bibinfo
  {author} {\bibfnamefont {E.}~\bibnamefont {Demler}},\ }\href {\doibase
  10.1103/PhysRevB.82.235114} {\bibfield  {journal} {\bibinfo  {journal} {Phys.
  Rev. B}\ }\textbf {\bibinfo {volume} {82}},\ \bibinfo {pages} {235114}
  (\bibinfo {year} {2010})}\BibitemShut {NoStop}%
\bibitem [{\citenamefont {Rudner}\ \emph {et~al.}(2013)\citenamefont {Rudner},
  \citenamefont {Lindner}, \citenamefont {Berg},\ and\ \citenamefont
  {Levin}}]{Rudner:prx13}%
  \BibitemOpen
  \bibfield  {author} {\bibinfo {author} {\bibfnamefont {M.~S.}\ \bibnamefont
  {Rudner}}, \bibinfo {author} {\bibfnamefont {N.~H.}\ \bibnamefont {Lindner}},
  \bibinfo {author} {\bibfnamefont {E.}~\bibnamefont {Berg}}, \ and\ \bibinfo
  {author} {\bibfnamefont {M.}~\bibnamefont {Levin}},\ }\href {\doibase
  10.1103/PhysRevX.3.031005} {\bibfield  {journal} {\bibinfo  {journal} {Phys.
  Rev. X}\ }\textbf {\bibinfo {volume} {3}},\ \bibinfo {pages} {031005}
  (\bibinfo {year} {2013})}\BibitemShut {NoStop}%
\bibitem [{\citenamefont {Katan}\ and\ \citenamefont
  {Podolsky}(2013)}]{Katan:prl13}%
  \BibitemOpen
  \bibfield  {author} {\bibinfo {author} {\bibfnamefont {Y.~T.}\ \bibnamefont
  {Katan}}\ and\ \bibinfo {author} {\bibfnamefont {D.}~\bibnamefont
  {Podolsky}},\ }\href {\doibase 10.1103/PhysRevLett.110.016802} {\bibfield
  {journal} {\bibinfo  {journal} {Phys. Rev. Lett.}\ }\textbf {\bibinfo
  {volume} {110}},\ \bibinfo {pages} {016802} (\bibinfo {year}
  {2013})}\BibitemShut {NoStop}%
\bibitem [{\citenamefont {Lindner}\ \emph {et~al.}(2013)\citenamefont
  {Lindner}, \citenamefont {Bergman}, \citenamefont {Refael},\ and\
  \citenamefont {Galitski}}]{Lindner:prb13}%
  \BibitemOpen
  \bibfield  {author} {\bibinfo {author} {\bibfnamefont {N.~H.}\ \bibnamefont
  {Lindner}}, \bibinfo {author} {\bibfnamefont {D.~L.}\ \bibnamefont
  {Bergman}}, \bibinfo {author} {\bibfnamefont {G.}~\bibnamefont {Refael}}, \
  and\ \bibinfo {author} {\bibfnamefont {V.}~\bibnamefont {Galitski}},\ }\href
  {\doibase 10.1103/PhysRevB.87.235131} {\bibfield  {journal} {\bibinfo
  {journal} {Phys. Rev. B}\ }\textbf {\bibinfo {volume} {87}},\ \bibinfo
  {pages} {235131} (\bibinfo {year} {2013})}\BibitemShut {NoStop}%
\bibitem [{\citenamefont {D\'ora}\ \emph {et~al.}(2012)\citenamefont {D\'ora},
  \citenamefont {Cayssol}, \citenamefont {Simon},\ and\ \citenamefont
  {Moessner}}]{Dora:prl12}%
  \BibitemOpen
  \bibfield  {author} {\bibinfo {author} {\bibfnamefont {B.}~\bibnamefont
  {D\'ora}}, \bibinfo {author} {\bibfnamefont {J.}~\bibnamefont {Cayssol}},
  \bibinfo {author} {\bibfnamefont {F.}~\bibnamefont {Simon}}, \ and\ \bibinfo
  {author} {\bibfnamefont {R.}~\bibnamefont {Moessner}},\ }\href {\doibase
  10.1103/PhysRevLett.108.056602} {\bibfield  {journal} {\bibinfo  {journal}
  {Phys. Rev. Lett.}\ }\textbf {\bibinfo {volume} {108}},\ \bibinfo {pages}
  {056602} (\bibinfo {year} {2012})}\BibitemShut {NoStop}%
\bibitem [{\citenamefont {Inoue}\ and\ \citenamefont
  {Tanaka}(2010)}]{Inoue:prl10}%
  \BibitemOpen
  \bibfield  {author} {\bibinfo {author} {\bibfnamefont {J.-i.}\ \bibnamefont
  {Inoue}}\ and\ \bibinfo {author} {\bibfnamefont {A.}~\bibnamefont {Tanaka}},\
  }\href {\doibase 10.1103/PhysRevLett.105.017401} {\bibfield  {journal}
  {\bibinfo  {journal} {Phys. Rev. Lett.}\ }\textbf {\bibinfo {volume} {105}},\
  \bibinfo {pages} {017401} (\bibinfo {year} {2010})}\BibitemShut {NoStop}%
\bibitem [{\citenamefont {Cayssol}\ \emph {et~al.}(2013)\citenamefont
  {Cayssol}, \citenamefont {Dora}, \citenamefont {Simon},\ and\ \citenamefont
  {Moessner}}]{Cayssol:pss13}%
  \BibitemOpen
  \bibfield  {author} {\bibinfo {author} {\bibfnamefont {J.}~\bibnamefont
  {Cayssol}}, \bibinfo {author} {\bibfnamefont {B.}~\bibnamefont {Dora}},
  \bibinfo {author} {\bibfnamefont {F.}~\bibnamefont {Simon}}, \ and\ \bibinfo
  {author} {\bibfnamefont {R.}~\bibnamefont {Moessner}},\ }\href {\doibase
  10.1002/pssr.201206451} {\bibfield  {journal} {\bibinfo  {journal} {physica
  status solidi (RRL) âRapid Research Letters}\ }\textbf {\bibinfo {volume}
  {7}},\ \bibinfo {pages} {101} (\bibinfo {year} {2013})}\BibitemShut {NoStop}%
\bibitem [{\citenamefont {Kitagawa}\ \emph {et~al.}(2011)\citenamefont
  {Kitagawa}, \citenamefont {Oka}, \citenamefont {Brataas}, \citenamefont
  {Fu},\ and\ \citenamefont {Demler}}]{Kitagawa:prb11}%
  \BibitemOpen
  \bibfield  {author} {\bibinfo {author} {\bibfnamefont {T.}~\bibnamefont
  {Kitagawa}}, \bibinfo {author} {\bibfnamefont {T.}~\bibnamefont {Oka}},
  \bibinfo {author} {\bibfnamefont {A.}~\bibnamefont {Brataas}}, \bibinfo
  {author} {\bibfnamefont {L.}~\bibnamefont {Fu}}, \ and\ \bibinfo {author}
  {\bibfnamefont {E.}~\bibnamefont {Demler}},\ }\href {\doibase
  10.1103/PhysRevB.84.235108} {\bibfield  {journal} {\bibinfo  {journal} {Phys.
  Rev. B}\ }\textbf {\bibinfo {volume} {84}},\ \bibinfo {pages} {235108}
  (\bibinfo {year} {2011})}\BibitemShut {NoStop}%
\bibitem [{\citenamefont {Iadecola}\ \emph {et~al.}(2013)\citenamefont
  {Iadecola}, \citenamefont {Campbell}, \citenamefont {Chamon}, \citenamefont
  {Hou}, \citenamefont {Jackiw}, \citenamefont {Pi},\ and\ \citenamefont
  {Kusminskiy}}]{Iadecola:prl13}%
  \BibitemOpen
  \bibfield  {author} {\bibinfo {author} {\bibfnamefont {T.}~\bibnamefont
  {Iadecola}}, \bibinfo {author} {\bibfnamefont {D.}~\bibnamefont {Campbell}},
  \bibinfo {author} {\bibfnamefont {C.}~\bibnamefont {Chamon}}, \bibinfo
  {author} {\bibfnamefont {C.-Y.}\ \bibnamefont {Hou}}, \bibinfo {author}
  {\bibfnamefont {R.}~\bibnamefont {Jackiw}}, \bibinfo {author} {\bibfnamefont
  {S.-Y.}\ \bibnamefont {Pi}}, \ and\ \bibinfo {author} {\bibfnamefont {S.~V.}\
  \bibnamefont {Kusminskiy}},\ }\href {\doibase 10.1103/PhysRevLett.110.176603}
  {\bibfield  {journal} {\bibinfo  {journal} {Phys. Rev. Lett.}\ }\textbf
  {\bibinfo {volume} {110}},\ \bibinfo {pages} {176603} (\bibinfo {year}
  {2013})}\BibitemShut {NoStop}%
\bibitem [{\citenamefont {Ezawa}(2013)}]{Ezawa:prl13}%
  \BibitemOpen
  \bibfield  {author} {\bibinfo {author} {\bibfnamefont {M.}~\bibnamefont
  {Ezawa}},\ }\href {\doibase 10.1103/PhysRevLett.110.026603} {\bibfield
  {journal} {\bibinfo  {journal} {Phys. Rev. Lett.}\ }\textbf {\bibinfo
  {volume} {110}},\ \bibinfo {pages} {026603} (\bibinfo {year}
  {2013})}\BibitemShut {NoStop}%
\bibitem [{\citenamefont {Kemper}\ \emph {et~al.}(2013)\citenamefont {Kemper},
  \citenamefont {Sentef}, \citenamefont {Moritz}, \citenamefont {Kao},
  \citenamefont {Shen}, \citenamefont {Freericks},\ and\ \citenamefont
  {Devereaux}}]{Kemper:prb13}%
  \BibitemOpen
  \bibfield  {author} {\bibinfo {author} {\bibfnamefont {A.~F.}\ \bibnamefont
  {Kemper}}, \bibinfo {author} {\bibfnamefont {M.}~\bibnamefont {Sentef}},
  \bibinfo {author} {\bibfnamefont {B.}~\bibnamefont {Moritz}}, \bibinfo
  {author} {\bibfnamefont {C.~C.}\ \bibnamefont {Kao}}, \bibinfo {author}
  {\bibfnamefont {Z.~X.}\ \bibnamefont {Shen}}, \bibinfo {author}
  {\bibfnamefont {J.~K.}\ \bibnamefont {Freericks}}, \ and\ \bibinfo {author}
  {\bibfnamefont {T.~P.}\ \bibnamefont {Devereaux}},\ }\href {\doibase
  10.1103/PhysRevB.87.235139} {\bibfield  {journal} {\bibinfo  {journal} {Phys.
  Rev. B}\ }\textbf {\bibinfo {volume} {87}},\ \bibinfo {pages} {235139}
  (\bibinfo {year} {2013})}\BibitemShut {NoStop}%
\bibitem [{\citenamefont {Rechtsman}\ \emph {et~al.}(2013)\citenamefont
  {Rechtsman}, \citenamefont {Zeuner}, \citenamefont {Plotnik}, \citenamefont
  {Lumer}, \citenamefont {Podolsky}, \citenamefont {Dreisow}, \citenamefont
  {Nolte}, \citenamefont {Segev},\ and\ \citenamefont
  {Szameit}}]{Rechtsman:nat13}%
  \BibitemOpen
  \bibfield  {author} {\bibinfo {author} {\bibfnamefont {M.~C.}\ \bibnamefont
  {Rechtsman}}, \bibinfo {author} {\bibfnamefont {J.~M.}\ \bibnamefont
  {Zeuner}}, \bibinfo {author} {\bibfnamefont {Y.}~\bibnamefont {Plotnik}},
  \bibinfo {author} {\bibfnamefont {Y.}~\bibnamefont {Lumer}}, \bibinfo
  {author} {\bibfnamefont {D.}~\bibnamefont {Podolsky}}, \bibinfo {author}
  {\bibfnamefont {F.}~\bibnamefont {Dreisow}}, \bibinfo {author} {\bibfnamefont
  {S.}~\bibnamefont {Nolte}}, \bibinfo {author} {\bibfnamefont
  {M.}~\bibnamefont {Segev}}, \ and\ \bibinfo {author} {\bibfnamefont
  {A.}~\bibnamefont {Szameit}},\ }\href {http://dx.doi.org/10.1038/nature12066}
  {\bibfield  {journal} {\bibinfo  {journal} {Nature}\ }\textbf {\bibinfo
  {volume} {496}},\ \bibinfo {pages} {196} (\bibinfo {year} {2013})},\ \bibinfo
  {note} {letter}\BibitemShut {NoStop}%
\bibitem [{\citenamefont {Jotzu}\ \emph {et~al.}(2014)\citenamefont {Jotzu},
  \citenamefont {Messer}, \citenamefont {Desbuquois}, \citenamefont {Lebrat},
  \citenamefont {Uehlinger}, \citenamefont {Greif},\ and\ \citenamefont
  {Esslinger}}]{Jotzu:nat14}%
  \BibitemOpen
  \bibfield  {author} {\bibinfo {author} {\bibfnamefont {G.}~\bibnamefont
  {Jotzu}}, \bibinfo {author} {\bibfnamefont {M.}~\bibnamefont {Messer}},
  \bibinfo {author} {\bibfnamefont {R.}~\bibnamefont {Desbuquois}}, \bibinfo
  {author} {\bibfnamefont {M.}~\bibnamefont {Lebrat}}, \bibinfo {author}
  {\bibfnamefont {T.}~\bibnamefont {Uehlinger}}, \bibinfo {author}
  {\bibfnamefont {D.}~\bibnamefont {Greif}}, \ and\ \bibinfo {author}
  {\bibfnamefont {T.}~\bibnamefont {Esslinger}},\ }\href {\doibase
  10.1038/nature13915} {\bibfield  {journal} {\bibinfo  {journal} {Nature}\
  }\textbf {\bibinfo {volume} {93}},\ \bibinfo {pages} {237} (\bibinfo {year}
  {2014})}\BibitemShut {NoStop}%
\bibitem [{\citenamefont {Bilitewski}\ and\ \citenamefont
  {Cooper}(2015)}]{Bilitewski:pra15}%
  \BibitemOpen
  \bibfield  {author} {\bibinfo {author} {\bibfnamefont {T.}~\bibnamefont
  {Bilitewski}}\ and\ \bibinfo {author} {\bibfnamefont {N.~R.}\ \bibnamefont
  {Cooper}},\ }\href {\doibase 10.1103/PhysRevA.91.063611} {\bibfield
  {journal} {\bibinfo  {journal} {Phys. Rev. A}\ }\textbf {\bibinfo {volume}
  {91}},\ \bibinfo {pages} {063611} (\bibinfo {year} {2015})}\BibitemShut
  {NoStop}%
\bibitem [{\citenamefont {Fregoso}\ \emph {et~al.}(2013)\citenamefont
  {Fregoso}, \citenamefont {Wang}, \citenamefont {Gedik},\ and\ \citenamefont
  {Galitski}}]{Fregoso:prb13}%
  \BibitemOpen
  \bibfield  {author} {\bibinfo {author} {\bibfnamefont {B.~M.}\ \bibnamefont
  {Fregoso}}, \bibinfo {author} {\bibfnamefont {Y.~H.}\ \bibnamefont {Wang}},
  \bibinfo {author} {\bibfnamefont {N.}~\bibnamefont {Gedik}}, \ and\ \bibinfo
  {author} {\bibfnamefont {V.}~\bibnamefont {Galitski}},\ }\href {\doibase
  10.1103/PhysRevB.88.155129} {\bibfield  {journal} {\bibinfo  {journal} {Phys.
  Rev. B}\ }\textbf {\bibinfo {volume} {88}},\ \bibinfo {pages} {155129}
  (\bibinfo {year} {2013})}\BibitemShut {NoStop}%
\bibitem [{\citenamefont {Sentef}\ \emph {et~al.}(2015)\citenamefont {Sentef},
  \citenamefont {Claassen}, \citenamefont {Kemper}, \citenamefont {Moritz},
  \citenamefont {Oka}, \citenamefont {Freericks},\ and\ \citenamefont
  {Devereaux}}]{sentef2015theory}%
  \BibitemOpen
  \bibfield  {author} {\bibinfo {author} {\bibfnamefont {M.}~\bibnamefont
  {Sentef}}, \bibinfo {author} {\bibfnamefont {M.}~\bibnamefont {Claassen}},
  \bibinfo {author} {\bibfnamefont {A.}~\bibnamefont {Kemper}}, \bibinfo
  {author} {\bibfnamefont {B.}~\bibnamefont {Moritz}}, \bibinfo {author}
  {\bibfnamefont {T.}~\bibnamefont {Oka}}, \bibinfo {author} {\bibfnamefont
  {J.}~\bibnamefont {Freericks}}, \ and\ \bibinfo {author} {\bibfnamefont
  {T.}~\bibnamefont {Devereaux}},\ }\href@noop {} {\bibfield  {journal}
  {\bibinfo  {journal} {Nature communications}\ }\textbf {\bibinfo {volume}
  {6}} (\bibinfo {year} {2015})}\BibitemShut {NoStop}%
\bibitem [{\citenamefont {Wang}\ \emph {et~al.}(2013)\citenamefont {Wang},
  \citenamefont {Steinberg}, \citenamefont {Jarillo-Herrero},\ and\
  \citenamefont {Gedik}}]{Wang:sci13}%
  \BibitemOpen
  \bibfield  {author} {\bibinfo {author} {\bibfnamefont {Y.~H.}\ \bibnamefont
  {Wang}}, \bibinfo {author} {\bibfnamefont {H.}~\bibnamefont {Steinberg}},
  \bibinfo {author} {\bibfnamefont {P.}~\bibnamefont {Jarillo-Herrero}}, \ and\
  \bibinfo {author} {\bibfnamefont {N.}~\bibnamefont {Gedik}},\ }\href
  {\doibase 10.1126/science.1239834} {\bibfield  {journal} {\bibinfo  {journal}
  {Science}\ }\textbf {\bibinfo {volume} {342}},\ \bibinfo {pages} {453}
  (\bibinfo {year} {2013})}\BibitemShut {NoStop}%
\bibitem [{\citenamefont {Mahmood}\ \emph {et~al.}(2016)\citenamefont
  {Mahmood}, \citenamefont {Chan}, \citenamefont {Alpichshev}, \citenamefont
  {Gardner}, \citenamefont {Lee}, \citenamefont {Lee},\ and\ \citenamefont
  {Gedik}}]{Mahmood:np16}%
  \BibitemOpen
  \bibfield  {author} {\bibinfo {author} {\bibfnamefont {F.}~\bibnamefont
  {Mahmood}}, \bibinfo {author} {\bibfnamefont {C.-K.}\ \bibnamefont {Chan}},
  \bibinfo {author} {\bibfnamefont {Z.}~\bibnamefont {Alpichshev}}, \bibinfo
  {author} {\bibfnamefont {D.}~\bibnamefont {Gardner}}, \bibinfo {author}
  {\bibfnamefont {Y.}~\bibnamefont {Lee}}, \bibinfo {author} {\bibfnamefont
  {P.~A.}\ \bibnamefont {Lee}}, \ and\ \bibinfo {author} {\bibfnamefont
  {N.}~\bibnamefont {Gedik}},\ }\href {http://dx.doi.org/10.1038/nphys3609}
  {\bibfield  {journal} {\bibinfo  {journal} {Nat Phys}\ }\textbf {\bibinfo
  {volume} {12}},\ \bibinfo {pages} {306} (\bibinfo {year} {2016})},\ \bibinfo
  {note} {letter}\BibitemShut {NoStop}%
\bibitem [{\citenamefont {Kim}\ \emph {et~al.}(2014)\citenamefont {Kim},
  \citenamefont {Ikeda},\ and\ \citenamefont {Huse}}]{Kim:pre14}%
  \BibitemOpen
  \bibfield  {author} {\bibinfo {author} {\bibfnamefont {H.}~\bibnamefont
  {Kim}}, \bibinfo {author} {\bibfnamefont {T.~N.}\ \bibnamefont {Ikeda}}, \
  and\ \bibinfo {author} {\bibfnamefont {D.~A.}\ \bibnamefont {Huse}},\ }\href
  {\doibase 10.1103/PhysRevE.90.052105} {\bibfield  {journal} {\bibinfo
  {journal} {Phys. Rev. E}\ }\textbf {\bibinfo {volume} {90}},\ \bibinfo
  {pages} {052105} (\bibinfo {year} {2014})}\BibitemShut {NoStop}%
\bibitem [{\citenamefont {Ponte}\ \emph {et~al.}(2015)\citenamefont {Ponte},
  \citenamefont {Papi\ifmmode~\acute{c}\else \'{c}\fi{}}, \citenamefont
  {Huveneers},\ and\ \citenamefont {Abanin}}]{Ponte:prl15}%
  \BibitemOpen
  \bibfield  {author} {\bibinfo {author} {\bibfnamefont {P.}~\bibnamefont
  {Ponte}}, \bibinfo {author} {\bibfnamefont {Z.}~\bibnamefont
  {Papi\ifmmode~\acute{c}\else \'{c}\fi{}}}, \bibinfo {author} {\bibfnamefont
  {F.~m.~c.}\ \bibnamefont {Huveneers}}, \ and\ \bibinfo {author}
  {\bibfnamefont {D.~A.}\ \bibnamefont {Abanin}},\ }\href {\doibase
  10.1103/PhysRevLett.114.140401} {\bibfield  {journal} {\bibinfo  {journal}
  {Phys. Rev. Lett.}\ }\textbf {\bibinfo {volume} {114}},\ \bibinfo {pages}
  {140401} (\bibinfo {year} {2015})}\BibitemShut {NoStop}%
\bibitem [{\citenamefont {Lazarides}\ \emph {et~al.}(2015)\citenamefont
  {Lazarides}, \citenamefont {Das},\ and\ \citenamefont
  {Moessner}}]{Lazarides:prl15}%
  \BibitemOpen
  \bibfield  {author} {\bibinfo {author} {\bibfnamefont {A.}~\bibnamefont
  {Lazarides}}, \bibinfo {author} {\bibfnamefont {A.}~\bibnamefont {Das}}, \
  and\ \bibinfo {author} {\bibfnamefont {R.}~\bibnamefont {Moessner}},\ }\href
  {\doibase 10.1103/PhysRevLett.115.030402} {\bibfield  {journal} {\bibinfo
  {journal} {Phys. Rev. Lett.}\ }\textbf {\bibinfo {volume} {115}},\ \bibinfo
  {pages} {030402} (\bibinfo {year} {2015})}\BibitemShut {NoStop}%
\bibitem [{\citenamefont {Genske}\ and\ \citenamefont
  {Rosch}(2015)}]{Genske:pra15}%
  \BibitemOpen
  \bibfield  {author} {\bibinfo {author} {\bibfnamefont {M.}~\bibnamefont
  {Genske}}\ and\ \bibinfo {author} {\bibfnamefont {A.}~\bibnamefont {Rosch}},\
  }\href {\doibase 10.1103/PhysRevA.92.062108} {\bibfield  {journal} {\bibinfo
  {journal} {Phys. Rev. A}\ }\textbf {\bibinfo {volume} {92}},\ \bibinfo
  {pages} {062108} (\bibinfo {year} {2015})}\BibitemShut {NoStop}%
\bibitem [{\citenamefont {Dehghani}\ \emph {et~al.}(2014)\citenamefont
  {Dehghani}, \citenamefont {Oka},\ and\ \citenamefont
  {Mitra}}]{Hossein:prb14}%
  \BibitemOpen
  \bibfield  {author} {\bibinfo {author} {\bibfnamefont {H.}~\bibnamefont
  {Dehghani}}, \bibinfo {author} {\bibfnamefont {T.}~\bibnamefont {Oka}}, \
  and\ \bibinfo {author} {\bibfnamefont {A.}~\bibnamefont {Mitra}},\ }\href
  {\doibase 10.1103/PhysRevB.90.195429} {\bibfield  {journal} {\bibinfo
  {journal} {Phys. Rev. B}\ }\textbf {\bibinfo {volume} {90}},\ \bibinfo
  {pages} {195429} (\bibinfo {year} {2014})}\BibitemShut {NoStop}%
\bibitem [{\citenamefont {Dehghani}\ and\ \citenamefont
  {Mitra}(2016{\natexlab{a}})}]{Hossein_2:prb16}%
  \BibitemOpen
  \bibfield  {author} {\bibinfo {author} {\bibfnamefont {H.}~\bibnamefont
  {Dehghani}}\ and\ \bibinfo {author} {\bibfnamefont {A.}~\bibnamefont
  {Mitra}},\ }\href {\doibase 10.1103/PhysRevB.93.245416} {\bibfield  {journal}
  {\bibinfo  {journal} {Phys. Rev. B}\ }\textbf {\bibinfo {volume} {93}},\
  \bibinfo {pages} {245416} (\bibinfo {year} {2016}{\natexlab{a}})}\BibitemShut
  {NoStop}%
\bibitem [{\citenamefont {Iadecola}\ and\ \citenamefont
  {Chamon}(2015)}]{Iadecola:prb15}%
  \BibitemOpen
  \bibfield  {author} {\bibinfo {author} {\bibfnamefont {T.}~\bibnamefont
  {Iadecola}}\ and\ \bibinfo {author} {\bibfnamefont {C.}~\bibnamefont
  {Chamon}},\ }\href {\doibase 10.1103/PhysRevB.91.184301} {\bibfield
  {journal} {\bibinfo  {journal} {Phys. Rev. B}\ }\textbf {\bibinfo {volume}
  {91}},\ \bibinfo {pages} {184301} (\bibinfo {year} {2015})}\BibitemShut
  {NoStop}%
\bibitem [{\citenamefont {Iadecola}\ \emph {et~al.}(2015)\citenamefont
  {Iadecola}, \citenamefont {Neupert},\ and\ \citenamefont
  {Chamon}}]{Iadecola_2:prb15}%
  \BibitemOpen
  \bibfield  {author} {\bibinfo {author} {\bibfnamefont {T.}~\bibnamefont
  {Iadecola}}, \bibinfo {author} {\bibfnamefont {T.}~\bibnamefont {Neupert}}, \
  and\ \bibinfo {author} {\bibfnamefont {C.}~\bibnamefont {Chamon}},\ }\href
  {\doibase 10.1103/PhysRevB.91.235133} {\bibfield  {journal} {\bibinfo
  {journal} {Phys. Rev. B}\ }\textbf {\bibinfo {volume} {91}},\ \bibinfo
  {pages} {235133} (\bibinfo {year} {2015})}\BibitemShut {NoStop}%
\bibitem [{\citenamefont {Seetharam}\ \emph {et~al.}(2015)\citenamefont
  {Seetharam}, \citenamefont {Bardyn}, \citenamefont {Lindner}, \citenamefont
  {Rudner},\ and\ \citenamefont {Refael}}]{Seetharam:prx15}%
  \BibitemOpen
  \bibfield  {author} {\bibinfo {author} {\bibfnamefont {K.~I.}\ \bibnamefont
  {Seetharam}}, \bibinfo {author} {\bibfnamefont {C.-E.}\ \bibnamefont
  {Bardyn}}, \bibinfo {author} {\bibfnamefont {N.~H.}\ \bibnamefont {Lindner}},
  \bibinfo {author} {\bibfnamefont {M.~S.}\ \bibnamefont {Rudner}}, \ and\
  \bibinfo {author} {\bibfnamefont {G.}~\bibnamefont {Refael}},\ }\href
  {\doibase 10.1103/PhysRevX.5.041050} {\bibfield  {journal} {\bibinfo
  {journal} {Phys. Rev. X}\ }\textbf {\bibinfo {volume} {5}},\ \bibinfo {pages}
  {041050} (\bibinfo {year} {2015})}\BibitemShut {NoStop}%
\bibitem [{\citenamefont {Shirai}\ \emph {et~al.}(2015)\citenamefont {Shirai},
  \citenamefont {Mori},\ and\ \citenamefont {Miyashita}}]{Shirai:pre15}%
  \BibitemOpen
  \bibfield  {author} {\bibinfo {author} {\bibfnamefont {T.}~\bibnamefont
  {Shirai}}, \bibinfo {author} {\bibfnamefont {T.}~\bibnamefont {Mori}}, \ and\
  \bibinfo {author} {\bibfnamefont {S.}~\bibnamefont {Miyashita}},\ }\href
  {\doibase 10.1103/PhysRevE.91.030101} {\bibfield  {journal} {\bibinfo
  {journal} {Phys. Rev. E}\ }\textbf {\bibinfo {volume} {91}},\ \bibinfo
  {pages} {030101} (\bibinfo {year} {2015})}\BibitemShut {NoStop}%
\bibitem [{\citenamefont {Bukov}\ \emph {et~al.}(2015)\citenamefont {Bukov},
  \citenamefont {Gopalakrishnan}, \citenamefont {Knap},\ and\ \citenamefont
  {Demler}}]{Bukov:prl15}%
  \BibitemOpen
  \bibfield  {author} {\bibinfo {author} {\bibfnamefont {M.}~\bibnamefont
  {Bukov}}, \bibinfo {author} {\bibfnamefont {S.}~\bibnamefont
  {Gopalakrishnan}}, \bibinfo {author} {\bibfnamefont {M.}~\bibnamefont
  {Knap}}, \ and\ \bibinfo {author} {\bibfnamefont {E.}~\bibnamefont
  {Demler}},\ }\href {\doibase 10.1103/PhysRevLett.115.205301} {\bibfield
  {journal} {\bibinfo  {journal} {Phys. Rev. Lett.}\ }\textbf {\bibinfo
  {volume} {115}},\ \bibinfo {pages} {205301} (\bibinfo {year}
  {2015})}\BibitemShut {NoStop}%
\bibitem [{\citenamefont {Dehghani}\ \emph {et~al.}(2015)\citenamefont
  {Dehghani}, \citenamefont {Oka},\ and\ \citenamefont
  {Mitra}}]{Aditiprb91-2015}%
  \BibitemOpen
  \bibfield  {author} {\bibinfo {author} {\bibfnamefont {H.}~\bibnamefont
  {Dehghani}}, \bibinfo {author} {\bibfnamefont {T.}~\bibnamefont {Oka}}, \
  and\ \bibinfo {author} {\bibfnamefont {A.}~\bibnamefont {Mitra}},\ }\href
  {\doibase 10.1103/PhysRevB.91.155422} {\bibfield  {journal} {\bibinfo
  {journal} {Phys. Rev. B}\ }\textbf {\bibinfo {volume} {91}},\ \bibinfo
  {pages} {155422} (\bibinfo {year} {2015})}\BibitemShut {NoStop}%
\bibitem [{\citenamefont {Dehghani}\ and\ \citenamefont
  {Mitra}(2015)}]{Aditiprb92-2015}%
  \BibitemOpen
  \bibfield  {author} {\bibinfo {author} {\bibfnamefont {H.}~\bibnamefont
  {Dehghani}}\ and\ \bibinfo {author} {\bibfnamefont {A.}~\bibnamefont
  {Mitra}},\ }\href {\doibase 10.1103/PhysRevB.92.165111} {\bibfield  {journal}
  {\bibinfo  {journal} {Phys. Rev. B}\ }\textbf {\bibinfo {volume} {92}},\
  \bibinfo {pages} {165111} (\bibinfo {year} {2015})}\BibitemShut {NoStop}%
\bibitem [{\citenamefont {Dehghani}\ and\ \citenamefont
  {Mitra}(2016{\natexlab{b}})}]{Hossein:prb16}%
  \BibitemOpen
  \bibfield  {author} {\bibinfo {author} {\bibfnamefont {H.}~\bibnamefont
  {Dehghani}}\ and\ \bibinfo {author} {\bibfnamefont {A.}~\bibnamefont
  {Mitra}},\ }\href {\doibase 10.1103/PhysRevB.93.205437} {\bibfield  {journal}
  {\bibinfo  {journal} {Phys. Rev. B}\ }\textbf {\bibinfo {volume} {93}},\
  \bibinfo {pages} {205437} (\bibinfo {year} {2016}{\natexlab{b}})}\BibitemShut
  {NoStop}%
\bibitem [{\citenamefont {Foa~Torres}\ \emph {et~al.}(2014)\citenamefont
  {Foa~Torres}, \citenamefont {Perez-Piskunow}, \citenamefont {Balseiro},\ and\
  \citenamefont {Usaj}}]{Torres:prl14}%
  \BibitemOpen
  \bibfield  {author} {\bibinfo {author} {\bibfnamefont {L.~E.~F.}\
  \bibnamefont {Foa~Torres}}, \bibinfo {author} {\bibfnamefont {P.~M.}\
  \bibnamefont {Perez-Piskunow}}, \bibinfo {author} {\bibfnamefont {C.~A.}\
  \bibnamefont {Balseiro}}, \ and\ \bibinfo {author} {\bibfnamefont
  {G.}~\bibnamefont {Usaj}},\ }\href {\doibase 10.1103/PhysRevLett.113.266801}
  {\bibfield  {journal} {\bibinfo  {journal} {Phys. Rev. Lett.}\ }\textbf
  {\bibinfo {volume} {113}},\ \bibinfo {pages} {266801} (\bibinfo {year}
  {2014})}\BibitemShut {NoStop}%
\bibitem{PhysRevB.91.241404}
H.~L. Calvo, L.~E.~F. Foa~Torres, P.~M. Perez-Piskunow, C.~A. Balseiro, and
  Gonzalo Usaj.
\newblock {\em Phys. Rev. B}, 91:241404, Jun 2015.

\bibitem{PhysRevA.92.023624}
V.~Dal~Lago, M.~Atala, and L.~E.~F. Foa~Torres.
\newblock {\em Phys. Rev. A}, 92:023624, Aug 2015.

\bibitem{PhysRevA.91.043625}
P.~M. Perez-Piskunow, L.~E.~F. Foa~Torres, and Gonzalo Usaj.
\newblock {\em Phys. Rev. A}, 91:043625, Apr 2015.

\bibitem{PhysRevB.89.121401}
P.~M. Perez-Piskunow, Gonzalo Usaj, C.~A. Balseiro, and L.~E. F.~Foa Torres.
\newblock {\em Phys. Rev. B}, 89:121401, Mar 2014.

\bibitem [{\citenamefont {Kundu}\ \emph {et~al.}(2014)\citenamefont {Kundu},
  \citenamefont {Fertig},\ and\ \citenamefont {Seradjeh}}]{Kundu:prl14}%
  \BibitemOpen
  \bibfield  {author} {\bibinfo {author} {\bibfnamefont {A.}~\bibnamefont
  {Kundu}}, \bibinfo {author} {\bibfnamefont {H.~A.}\ \bibnamefont {Fertig}}, \
  and\ \bibinfo {author} {\bibfnamefont {B.}~\bibnamefont {Seradjeh}},\ }\href
  {\doibase 10.1103/PhysRevLett.113.236803} {\bibfield  {journal} {\bibinfo
  {journal} {Phys. Rev. Lett.}\ }\textbf {\bibinfo {volume} {113}},\ \bibinfo
  {pages} {236803} (\bibinfo {year} {2014})}\BibitemShut {NoStop}%
\bibitem [{\citenamefont {Farrell}\ and\ \citenamefont
  {Pereg-Barnea}(2016)}]{Farrell:prb16}%
  \BibitemOpen
  \bibfield  {author} {\bibinfo {author} {\bibfnamefont {A.}~\bibnamefont
  {Farrell}}\ and\ \bibinfo {author} {\bibfnamefont {T.}~\bibnamefont
  {Pereg-Barnea}},\ }\href {\doibase 10.1103/PhysRevB.93.045121} {\bibfield
  {journal} {\bibinfo  {journal} {Phys. Rev. B}\ }\textbf {\bibinfo {volume}
  {93}},\ \bibinfo {pages} {045121} (\bibinfo {year} {2016})}\BibitemShut
  {NoStop}%
\bibitem [{\citenamefont {D?Alessio}\ and\ \citenamefont
  {Polkovnikov}(2013)}]{D?Alessio201319}%
  \BibitemOpen
  \bibfield  {author} {\bibinfo {author} {\bibfnamefont {L.}~\bibnamefont
  {D'Alessio}}\ and\ \bibinfo {author} {\bibfnamefont {A.}~\bibnamefont
  {Polkovnikov}},\ }\href {\doibase
  http://dx.doi.org/10.1016/j.aop.2013.02.011} {\bibfield  {journal} {\bibinfo
  {journal} {Annals of Physics}\ }\textbf {\bibinfo {volume} {333}},\ \bibinfo
  {pages} {19 } (\bibinfo {year} {2013})}\BibitemShut {NoStop}%
\bibitem [{\citenamefont {Grushin}\ \emph {et~al.}(2014)\citenamefont
  {Grushin}, \citenamefont {G\'omez-Le\'on},\ and\ \citenamefont
  {Neupert}}]{Grushin:prl14}%
  \BibitemOpen
  \bibfield  {author} {\bibinfo {author} {\bibfnamefont {A.~G.}\ \bibnamefont
  {Grushin}}, \bibinfo {author} {\bibfnamefont {A.}~\bibnamefont
  {G\'omez-Le\'on}}, \ and\ \bibinfo {author} {\bibfnamefont {T.}~\bibnamefont
  {Neupert}},\ }\href {\doibase 10.1103/PhysRevLett.112.156801} {\bibfield
  {journal} {\bibinfo  {journal} {Phys. Rev. Lett.}\ }\textbf {\bibinfo
  {volume} {112}},\ \bibinfo {pages} {156801} (\bibinfo {year}
  {2014})}\BibitemShut {NoStop}%
\bibitem [{\citenamefont {Guo}\ and\ \citenamefont {Franz}(2009)}]{Guo:prb09}%
  \BibitemOpen
  \bibfield  {author} {\bibinfo {author} {\bibfnamefont {H.-M.}\ \bibnamefont
  {Guo}}\ and\ \bibinfo {author} {\bibfnamefont {M.}~\bibnamefont {Franz}},\
  }\href {\doibase 10.1103/PhysRevB.80.113102} {\bibfield  {journal} {\bibinfo
  {journal} {Phys. Rev. B}\ }\textbf {\bibinfo {volume} {80}},\ \bibinfo {eid}
  {113102} (\bibinfo {year} {2009})}\BibitemShut {NoStop}%
\bibitem [{\citenamefont {Hermele}\ \emph {et~al.}(2008)\citenamefont
  {Hermele}, \citenamefont {Ran}, \citenamefont {Lee},\ and\ \citenamefont
  {Wen}}]{Hermele-prb77-2008}%
  \BibitemOpen
  \bibfield  {author} {\bibinfo {author} {\bibfnamefont {M.}~\bibnamefont
  {Hermele}}, \bibinfo {author} {\bibfnamefont {Y.}~\bibnamefont {Ran}},
  \bibinfo {author} {\bibfnamefont {P.~A.}\ \bibnamefont {Lee}}, \ and\
  \bibinfo {author} {\bibfnamefont {X.-G.}\ \bibnamefont {Wen}},\ }\href
  {\doibase 10.1103/PhysRevB.77.224413} {\bibfield  {journal} {\bibinfo
  {journal} {Phys. Rev. B}\ }\textbf {\bibinfo {volume} {77}},\ \bibinfo
  {pages} {224413} (\bibinfo {year} {2008})}\BibitemShut {NoStop}%
\bibitem [{\citenamefont {Floquet}(1883)}]{Floquet:1883}%
  \BibitemOpen
  \bibfield  {author} {\bibinfo {author} {\bibfnamefont {G.}~\bibnamefont
  {Floquet}},\ }\href@noop {} {\bibfield  {journal} {\bibinfo  {journal} {Ann.
  Sci. Ec. Normale Super.}\ }\textbf {\bibinfo {volume} {12}},\ \bibinfo
  {pages} {47} (\bibinfo {year} {1883})}\BibitemShut {NoStop}%
\bibitem [{\citenamefont {Haldane}(1988)}]{Haldane-prl61-1988}%
  \BibitemOpen
  \bibfield  {author} {\bibinfo {author} {\bibfnamefont {F.~D.~M.}\
  \bibnamefont {Haldane}},\ }\href {\doibase 10.1103/PhysRevLett.61.2015}
  {\bibfield  {journal} {\bibinfo  {journal} {Phys. Rev. Lett.}\ }\textbf
  {\bibinfo {volume} {61}},\ \bibinfo {pages} {2015} (\bibinfo {year}
  {1988})}\BibitemShut {NoStop}%
\bibitem [{\citenamefont {Polkovnikov}\ \emph {et~al.}(2011)\citenamefont
  {Polkovnikov}, \citenamefont {Sengupta}, \citenamefont {Silva},\ and\
  \citenamefont {Vengalattore}}]{Polkovnikov:rmp11}%
  \BibitemOpen
  \bibfield  {author} {\bibinfo {author} {\bibfnamefont {A.}~\bibnamefont
  {Polkovnikov}}, \bibinfo {author} {\bibfnamefont {K.}~\bibnamefont
  {Sengupta}}, \bibinfo {author} {\bibfnamefont {A.}~\bibnamefont {Silva}}, \
  and\ \bibinfo {author} {\bibfnamefont {M.}~\bibnamefont {Vengalattore}},\
  }\href {\doibase 10.1103/RevModPhys.83.863} {\bibfield  {journal} {\bibinfo
  {journal} {Rev. Mod. Phys.}\ }\textbf {\bibinfo {volume} {83}},\ \bibinfo
  {pages} {863} (\bibinfo {year} {2011})}\BibitemShut {NoStop}%
\bibitem [{\citenamefont {Petersen}\ \emph {et~al.}(2011)\citenamefont
  {Petersen}, \citenamefont {Kaiser}, \citenamefont {Dean}, \citenamefont
  {Simoncig}, \citenamefont {Liu}, \citenamefont {Cavalieri}, \citenamefont
  {Cacho}, \citenamefont {Turcu}, \citenamefont {Springate}, \citenamefont
  {Frassetto}, \citenamefont {Poletto}, \citenamefont {Dhesi}, \citenamefont
  {Berger},\ and\ \citenamefont {Cavalleri}}]{Petersen:prl11}%
  \BibitemOpen
  \bibfield  {author} {\bibinfo {author} {\bibfnamefont {J.~C.}\ \bibnamefont
  {Petersen}}, \bibinfo {author} {\bibfnamefont {S.}~\bibnamefont {Kaiser}},
  \bibinfo {author} {\bibfnamefont {N.}~\bibnamefont {Dean}}, \bibinfo {author}
  {\bibfnamefont {A.}~\bibnamefont {Simoncig}}, \bibinfo {author}
  {\bibfnamefont {H.~Y.}\ \bibnamefont {Liu}}, \bibinfo {author} {\bibfnamefont
  {A.~L.}\ \bibnamefont {Cavalieri}}, \bibinfo {author} {\bibfnamefont
  {C.}~\bibnamefont {Cacho}}, \bibinfo {author} {\bibfnamefont {I.~C.~E.}\
  \bibnamefont {Turcu}}, \bibinfo {author} {\bibfnamefont {E.}~\bibnamefont
  {Springate}}, \bibinfo {author} {\bibfnamefont {F.}~\bibnamefont
  {Frassetto}}, \bibinfo {author} {\bibfnamefont {L.}~\bibnamefont {Poletto}},
  \bibinfo {author} {\bibfnamefont {S.~S.}\ \bibnamefont {Dhesi}}, \bibinfo
  {author} {\bibfnamefont {H.}~\bibnamefont {Berger}}, \ and\ \bibinfo {author}
  {\bibfnamefont {A.}~\bibnamefont {Cavalleri}},\ }\href {\doibase
  10.1103/PhysRevLett.107.177402} {\bibfield  {journal} {\bibinfo  {journal}
  {Phys. Rev. Lett.}\ }\textbf {\bibinfo {volume} {107}},\ \bibinfo {pages}
  {177402} (\bibinfo {year} {2011})}\BibitemShut {NoStop}%
\bibitem [{\citenamefont {Morimoto}\ \emph {et~al.}(2009)\citenamefont
  {Morimoto}, \citenamefont {Hatsugai},\ and\ \citenamefont
  {Aoki}}]{Morimoto:prl09}%
  \BibitemOpen
  \bibfield  {author} {\bibinfo {author} {\bibfnamefont {T.}~\bibnamefont
  {Morimoto}}, \bibinfo {author} {\bibfnamefont {Y.}~\bibnamefont {Hatsugai}},
  \ and\ \bibinfo {author} {\bibfnamefont {H.}~\bibnamefont {Aoki}},\ }\href
  {\doibase 10.1103/PhysRevLett.103.116803} {\bibfield  {journal} {\bibinfo
  {journal} {Phys. Rev. Lett.}\ }\textbf {\bibinfo {volume} {103}},\ \bibinfo
  {pages} {116803} (\bibinfo {year} {2009})}\BibitemShut {NoStop}%
\bibitem [{\citenamefont {O'Connell}\ and\ \citenamefont
  {Wallace}(1982)}]{O'Connell:prb82}%
  \BibitemOpen
  \bibfield  {author} {\bibinfo {author} {\bibfnamefont {R.~F.}\ \bibnamefont
  {O'Connell}}\ and\ \bibinfo {author} {\bibfnamefont {G.}~\bibnamefont
  {Wallace}},\ }\href {\doibase 10.1103/PhysRevB.26.2231} {\bibfield  {journal}
  {\bibinfo  {journal} {Phys. Rev. B}\ }\textbf {\bibinfo {volume} {26}},\
  \bibinfo {pages} {2231} (\bibinfo {year} {1982})}\BibitemShut {NoStop}%
\bibitem [{\citenamefont {Ikebe}\ \emph {et~al.}(2010)\citenamefont {Ikebe},
  \citenamefont {Morimoto}, \citenamefont {Masutomi}, \citenamefont {Okamoto},
  \citenamefont {Aoki},\ and\ \citenamefont {Shimano}}]{Ikebe:prl10}%
  \BibitemOpen
  \bibfield  {author} {\bibinfo {author} {\bibfnamefont {Y.}~\bibnamefont
  {Ikebe}}, \bibinfo {author} {\bibfnamefont {T.}~\bibnamefont {Morimoto}},
  \bibinfo {author} {\bibfnamefont {R.}~\bibnamefont {Masutomi}}, \bibinfo
  {author} {\bibfnamefont {T.}~\bibnamefont {Okamoto}}, \bibinfo {author}
  {\bibfnamefont {H.}~\bibnamefont {Aoki}}, \ and\ \bibinfo {author}
  {\bibfnamefont {R.}~\bibnamefont {Shimano}},\ }\href {\doibase
  10.1103/PhysRevLett.104.256802} {\bibfield  {journal} {\bibinfo  {journal}
  {Phys. Rev. Lett.}\ }\textbf {\bibinfo {volume} {104}},\ \bibinfo {pages}
  {256802} (\bibinfo {year} {2010})}\BibitemShut {NoStop}%
\bibitem [{\citenamefont {Crassee}\ \emph {et~al.}(2011)\citenamefont
  {Crassee}, \citenamefont {Levallois}, \citenamefont {Walter}, \citenamefont
  {Ostler}, \citenamefont {Bostwick}, \citenamefont {Rotenberg}, \citenamefont
  {Seyller}, \citenamefont {van~der Marel},\ and\ \citenamefont
  {Kuzmenko}}]{Crassee:np11}%
  \BibitemOpen
  \bibfield  {author} {\bibinfo {author} {\bibfnamefont {I.}~\bibnamefont
  {Crassee}}, \bibinfo {author} {\bibfnamefont {J.}~\bibnamefont {Levallois}},
  \bibinfo {author} {\bibfnamefont {A.~L.}\ \bibnamefont {Walter}}, \bibinfo
  {author} {\bibfnamefont {M.}~\bibnamefont {Ostler}}, \bibinfo {author}
  {\bibfnamefont {A.}~\bibnamefont {Bostwick}}, \bibinfo {author}
  {\bibfnamefont {E.}~\bibnamefont {Rotenberg}}, \bibinfo {author}
  {\bibfnamefont {T.}~\bibnamefont {Seyller}}, \bibinfo {author} {\bibfnamefont
  {D.}~\bibnamefont {van~der Marel}}, \ and\ \bibinfo {author} {\bibfnamefont
  {A.~B.}\ \bibnamefont {Kuzmenko}},\ }\href {\doibase 10.1038/nphys1816}
  {\bibfield  {journal} {\bibinfo  {journal} {Nat Phys}\ }\textbf {\bibinfo
  {volume} {7}},\ \bibinfo {pages} {48} (\bibinfo {year} {2011})}\BibitemShut
  {NoStop}%
\bibitem [{\citenamefont {Shimano}\ \emph {et~al.}(2013)\citenamefont
  {Shimano}, \citenamefont {Yumoto}, \citenamefont {Yoo}, \citenamefont
  {Matsunaga}, \citenamefont {Tanabe}, \citenamefont {Hibino}, \citenamefont
  {Morimoto},\ and\ \citenamefont {Aoki}}]{Shimano:nc13}%
  \BibitemOpen
  \bibfield  {author} {\bibinfo {author} {\bibfnamefont {R.}~\bibnamefont
  {Shimano}}, \bibinfo {author} {\bibfnamefont {G.}~\bibnamefont {Yumoto}},
  \bibinfo {author} {\bibfnamefont {J.~Y.}\ \bibnamefont {Yoo}}, \bibinfo
  {author} {\bibfnamefont {R.}~\bibnamefont {Matsunaga}}, \bibinfo {author}
  {\bibfnamefont {S.}~\bibnamefont {Tanabe}}, \bibinfo {author} {\bibfnamefont
  {H.}~\bibnamefont {Hibino}}, \bibinfo {author} {\bibfnamefont
  {T.}~\bibnamefont {Morimoto}}, \ and\ \bibinfo {author} {\bibfnamefont
  {H.}~\bibnamefont {Aoki}},\ }\href {http://dx.doi.org/10.1038/ncomms2866}
  {\bibfield  {journal} {\bibinfo  {journal} {Nat Commun}\ }\textbf {\bibinfo
  {volume} {4}},\ \bibinfo {pages} {1841} (\bibinfo {year} {2013})},\ \bibinfo
  {note} {article}\BibitemShut {NoStop}%
\end{thebibliography}
%\bibliographystyle{plain}
%\bibliographystyle{aipnum4-1}
%merlin.mbs apsrev4-1.bst 2010-07-25 4.21a (PWD, AO, DPC) hacked
%Control: key (0)
%Control: author (8) initials jnrlst
%Control: editor formatted (1) identically to author
%Control: production of article title (-1) disabled
%Control: page (0) single
%Control: year (1) truncated
%Control: production of eprint (0) enabled
%

\end{document}